\documentclass[useAMS,usenatbib,fleqn]{mn2e}
\pdfoutput=1
\usepackage{graphicx,epsfig}
\usepackage{amsmath}
\usepackage{booktabs}
\usepackage{tabularx}
\usepackage{mathrsfs}

\def\gsim{\;\lower4pt\hbox{${\buildrel\displaystyle >\over\sim}$}\;}
\def\lsim{\;\lower4pt\hbox{${\buildrel\displaystyle <\over\sim}$}\;}
\def\grls{\;\lower4pt\hbox{${\buildrel\displaystyle >\over <}  $}\;}

\title[Void Dynamics with Shocks in Various Envelopes]
{Dynamic Voids Surrounded by Shocked\\
  Conventional Polytropic Gas Envelopes}
\author[Yu-Qing Lou and Lile Wang]{Yu-Qing Lou$^{1,2,3}$\thanks{E-mail:
louyq@tsinghua.edu.cn (Y-QL) } and Lile
Wang$^{1}$\thanks{wll90@126.com (LLW) }
\\
$^1$Department of Physics and Tsinghua Centre for Astrophysics
 (THCA), Tsinghua University, Beijing 100084, China \\
$^2$Department of Astronomy and Astrophysics, the University
 of Chicago, 5640 S. Ellis Ave, Chicago, IL 60637, USA \\
$^3$National Astronomical Observatories, Chinese Academy
 of Sciences, A20, Datun Road, Beijing 100021, China }

\date{Accepted 2011 September 12. Received 2011 August 22;
    in original form 2011 August 22}
\pubyear{2011}

\pagerange{\pageref{firstpage}--\pageref{lastpage}}

\begin{document}
\maketitle

\label{firstpage}

\begin{abstract}

With proper physical mechanisms of energy and momentum input
 from around the centre of a self-gravitating polytropic gas
 sphere, a central spherical ``void" or ``cavity" or ``bubble"
 of very much less mass contents may emerge and then
 dynamically expand into a variety of surrounding more
 massive gas envelopes with or without shocks.
We explore self-similar evolution of a self-gravitating
 polytropic hydrodynamic flow of spherical symmetry with
 such an expanding ``void" embedded around the center.
The void boundary supporting a massive envelope represents a
 pressure-balanced contact discontinuity where drastic changes
 in mass density and temperature occur.
We obtain numerical void solutions that can cross the
  sonic critical surface either smoothly or by shocks.
Using the conventional polytropic equation of state,
%  ($p=\kappa\rho^\gamma$, where $p$ is the thermal pressure,
%  $\rho$ the mass density, $\gamma$ the polytropic index and
%  $\kappa$ a constant)
 we construct global void solutions with shocks travelling
 into various envelopes including static polytropic sphere,
 outflow, inflow, breeze and contraction types.
%which have to cross the sonic critical surface. Solutions
%can cross the critical surface smoothly or by shocks, and
%we have studied both cases. Numerical integration of
%self-similar hydrodynamic equations, with different
%parameters, are applied to give some specific profiles
%  of various kinds of solutions.
%Especially for those void solutions with shocks, we have attempted
%  to construct such solutions with different kind of envelope
%  \citep[EWCS, outflow, inflow, breeze and contraction, for
%  definitions see][] {2008MNRAS.384..611L}, figuring out that there is
%  a restriction on similarity parameter $n$ (radius $r\propto t^n$ in
%  self-similar evolution) for us to obtain specific types of solutions:
%  those with EWCS, breeze and contraction envelopes requires $n>0.80$.
%  As applications in real-world
In the context of supernovae, we discuss the possible scenario of
 separating a central collapsing compact object from an outgoing
 gas envelope with a powerful void in dynamic expansion.
Initially, a central bubble is carved out by an extremely
 powerful neutrinosphere.
After the escape of neutrinos during the decoupling, the strong
 electromagnetic radiation field and/or electron-positron pair
 plasma continue to drive the cavity expansion.
In a self-similar dynamic evolution, the pressure across the
 contact discontinuity decreases with time to a negligible
 level for a sufficiently long lapse and eventually, the
 gas envelope continues to expand by inertia.
We describe model cases of polytropic index $\gamma=4/3-\epsilon$
 with $\epsilon>0$ and discuss pertinent requirements to justify
 our proposed scenario.

%For astrophysical applications, we have investigated in realistic
%  context the bubbles inside which materials are extremely
%  relativistic (in statistical physics, this leads to
%  $\gamma=4/3-\epsilon$, $0<\epsilon<1/3$; bubbles with such
%  $\gamma$'s are usually referred to as ``hot bubbles''), while
%  capabilities of doing so are examined (e.g., diffusion effects
%  across the contact discontinuities are estimated to be not
%  significant).
%For supernova remnants, Wilson mechanisms of explosion
%  \citep{1985ApJ...295...14B, RevModPhys.62.801} have suggested that
%  there be a cavity in the center of the ejecta formed by intense
%  neutrino flux before neutrinos decouple from the gas shell.
%We discuss in this article that it could be the radiation field
% (photon gases together with electron-positron pairs, etc.) that
% fills the central void thereafter.
%We have discussed the possibility of the central radiation fields'
%  being dominant power driving the ejecta outwards after neutrinos'
%  decoupling.
%As an detailed example, we have applied those models to describe
%  the dynamic evolution of SNR's and showed the capability of doing so.
%A possible version of deceleration in expansion has been discussed
%  of its legitimacy and applicability.
%
%  {\bf Pay attention to the concept of hot bubble (e.g. Bethe and
%    references therein). }
%  {\it I have abandoned the previous version of abstract and this
%    version is a totally new one.}
\end{abstract}


\section{Introduction}
\label{sec:INTRODUCTION}

Voids of much less density as compared to more massive and
 grossly spherical surroundings are fairly common in
 astrophysical systems on various spatial and temporal scales.
They have been found in a wide diversity of settings, including
 planetary nebulae (PNe) \citep[e.g. NGC 7662 and NGC 40;]
 [Lou \& Zhai 2010]{2004AJ....128.1705G}, bubbles and superbubbles
 \citep[e.g.][]{1987ApJ...317..190M,1999A&A...350..230K}
 in the interstellar medium (ISM), supernovae (SNe),
 supernova remnants (SNRs) and so forth.
It is conceivable that with sustained powerful sources of
 energy and momentum released from around the central region
 of a self-gravitating gas sphere, a considerably rarified
 void or cavity or bubble can form and dynamically expand
 into a more massive surrounding envelope.
Depending astrophysical contexts, such sustained central
 sources could be tenuous stellar winds, magnetized
 relativistic pulsar winds of mainly electron-positron
 pair plasma, energetic neutrino flux, and radiation
 field of trapped photon gas etc.
We advance in this paper a theoretical model scenario, which
 is formulated within the framework of self-similar evolution
 for a conventional polytropic gas, towards general profiles
 of spherically symmetric hydrodynamic systems embedded with
 central voids in expansion.
As a simplifying approximation, these voids are treated as
 massless quasi-spherical cavities whose gravitational fields
 are negligible when considering the dynamic evolution of the
 gas shells surrounding the central voids.
Naturally, there are always some materials
 inside the boundary of any void in reality.
% , more or less.
They are even indispensable for explaining the dynamic
 evolution of the shell outside in aspects other than gravity
 (e.g., pressure driving, tenuous wind driving, extremely
 relativistic light particles, and photon gas etc.).
With these qualifications in mind, our analysis would
 indicate that this model is capable and applicable for
 describing a variety of polytropic gaseous astrophysical
 envelopes in hydrodynamic evolution.
Some similar analysis along this line can be found in
 \cite[][2010]{springerlink:10.1007/s10509-009-0044-4},
 which focus on isothermal cases (i.e. isothermal
 self-similar voids -- ISSV) and astrophysical
 applications to PNe NGC 40 and NGC 7662 as examples;
 such central ISSV in PNe are powered by hot tenuous
 stellar winds with shocks.

Originated from extremely violent processes like supernova
 explosions, ``hot bubbles" or voids filled with intense
 radiation fields are a significant kind of structures in
 the evolution of many astrophysical systems including
 supernova remnants (SNRs).
One of our main concerns here is on various void
 structures in the self-similar dynamic evolution
 of SNe and subsequent SNRs.
\citet{1999ApJ...510..379M} noted the existence of
 similarities in the evolution of SNe at a very early
 epoch -- much earlier than the Sedov stage known for
 its characteristics of self-similar evolution.
It would be a first approximation to investigate dynamic
 evolution of SNe, whose mass is sufficiently large
 ($\gsim 8M_\odot$) as compared with the mass of the
 central collapsed compact object ($\sim 1.5M_\odot$),
 by self-similar hydrodynamic models with central voids.

Formation of such initial ``voids" within one or two
 hundred kilometers is plausibly the result of intense
 neutrino flux heating and driving, leading to the
 emergence of a rebound shock.
This Wilson mechanism was put forward
 and elaborated by \citet{1985ApJ...295...14B}
 \citep[and subsequently in][]{RevModPhys.62.801},
 where the authors argued that acceleration of mass infall
 towards the center is inhibited by the neutrino pressure
 from the so-called neutrinosphere with little amount of mass.
This mechanism is highlighted in numerical simulations of
 \citet{1989A&A...224...49J} and \citet{1996A&A...306..167J}
 for examples.
These simulation works have sketched a picture by simulation
 that stellar materials are heated and pushed outwards by the
 intense neutrino flux generated by nuclear processes during
 the rapid core collapse at the centre.
As a by-product of rebound shock revitalization, a bubble or
 cavity will have already been shaped up around the center
 during the epoch before the surrounding gas materials
 become too tenuous ($\sim 10^9\mbox{ g }\mbox{cm}^{-3}$
 as discussed presently in subsection
 \ref{sec:MOMENTUM_TRANSFER_ESTIMATION})
 to trap extremely energetic neutrinos.
Then the envelope shell, consisting of the vast majority of
 stellar materials, is expected to be ejected by the revived
 rebound shock.
After a short while,
%However, this expansion maintains a
%dramatic energy consumption, while
 relativistic neutrinos and outer surrounding
 massive gas envelope becomes completely decoupled.

We further discuss this physical feasibility regarding
 two factors: energy and momentum transfers from photons
 to an ionized plasma are very effective (compared with
 that from neutrinos to gas);
the energy carried by radiative emissions of photons is
 of the same order of magnitude ($\sim 10^{51}\mbox{ erg}$)
 as the energy needed to blow a stellar envelope up.
%{\it In \citet{2007ApJ...668L..55C}, a mechanism of radiation
% triggered SNe explosions has been proposed and discussed, yet
% there was not too much serious and systematic discussions on
% the hydrodynamic models of such a mechanism beyond some basic
% estimates. }
Our dynamic void solutions show that it could be the photon
 radiation field trapped inside the central cavity that
 drives an outer shell to expand after the decoupling and
 escape of energetic neutrinos.
As simplifications, we would treat a void containing a
 uniform radiation field instead of a mixture of conventional
 matters and radiation coupled by transport equations.
The scale of a central void at the very early stage
 \citep[$\gsim 100\mbox{ km}$, e.g.][]{RevModPhys.62.801}
 makes this uniformity approximation plausible as the
 perturbation in radiation field travels at the speed
 of light $c$.
Here the concept of radiation field is usually generalized as
 a combination of photons (electromagnetic field) and various
 products of pair-production such as electron-positron pairs,
 since the temperatures are always sufficiently high (at least
 $k_\text{B}T>1\mbox{ MeV}$) for pair production, where $k_B$
 is the Boltzmann constant.
This radiation-driven envelope expansion with self-gravity
 is then responsible for the acceleration and/or deceleration
 in the expansion of a SN;
we shall come back to this possibility later in this paper.

We conduct in this paper a systematic analysis of the self-similar
 hydrodynamic evolution of void surrounded by spherical envelopes
 with various radial structures using the similarity transformation.
Self-similar hydrodynamics with spherical symmetry has
 been extensively studied as a valuable tool with various
 approximations and simplifications.
The research works of
 \citet{1969MNRAS.145..405L,1969MNRAS.145..271L},
 \citet{1969MNRAS.145..457P,1969MNRAS.144..425P},
 \citet{1977ApJ...218..834H,1986MNRAS.223..391H},
 \citet{1977ApJ...214..488S}, \citet{1995ApJ...448..774T},
 \citet{1997ApJ...488..263C}, \citet{2002ApJ...580..969S},
 \citet{2004MNRAS.348..717L}, \citet{2004ApJ...611L.117S}, and
 \citet{2005MNRAS.363.1315B}, have investigated self-similar
 hydrodynamics and their astrophysical applications
 with an isothermal equation of state (EoS).
Meanwhile, those with polytropic EoS have
 been studied by \citet{1980ApJ...238..991G},
 \citet{1983ApJ...265.1047Y}, \citet{1985ApJ...288..644L},
 \citet{1988ApJ...326..527S}, \citet{2006MNRAS.372..885L},
 \citet{2008MNRAS.390.1619H}, and \citet{2008MNRAS.384..611L}.
Some of them are more pertinent to our investigation here,
 e.g. \citet{1977ApJ...218..834H} obtained a systematic
 self-similar transformation and obtained the isothermal
 expansion-wave collapse solution (EWCS), which has been
 substantially generalized;
%as the foundation of many other works;
\citet{1995ApJ...448..774T} introduced a procedure
 of treating shocks in an isothermal gas;
\citet{1983ApJ...265.1047Y} and \citet{1988ApJ...326..527S}
 developed self-similar procedure for conventional
 polytropic gas dynamics; \citet{2006MNRAS.372..885L} and
 \citet{2008MNRAS.384..611L} explored the scheme for the evolution
 of conventional and general polytropic gases with shocks.
Self-similar transformations permit a series of central-void
 solutions (\citealt[][2010]{springerlink:10.1007/s10509-009-0044-4} for
 isothermal cases and \citealt{2008MNRAS.390.1619H} for polytropic
 voids as examples) whose boundaries appear to be very steep in
%   sharp ``cliffs'', or {\bf precipitations?} of
 mass density.
This treatment is again an idealization and the justification
 of doing so is discussed in Appendix \ref {sec:DIFFUSION}.
We are also interested in the construction of self-similar
 conventional polytropic shock solutions with expanding
 central voids.

We note \citet{1983ApJ...265.1047Y} and
 \citet{2008MNRAS.384..611L} for their analyses of cases
 whose polytropic indices are $\gamma\rightarrow (4/3)^+$
 and $\gamma=4/3$.
For either relativistically hot or degenerate materials, it
 is a very good approximation to describe their behaviour
 by an adiabatic EoS $p=\kappa\rho^{4/3}$.
Here $\kappa$ is a function of $(r,\ t)$ in general polytropic
 cases satisfying specific entropy conservation along
 streamlines \citep[e.g.][]{2004ApJ...615..813F}.
Specification of $\kappa$ as a global constant is a
 special case, i.e., a conventional polytropic gas.
% (might be different across discontinuities like shocks).
An exact $\gamma =4/3$ case is analyzed by
 \citet{1980ApJ...238..991G} for a homologous
 core collapse solution.
In this paper, we emphasize void solutions with
 conventional polytropic conditions whose $\gamma$
 takes the form of $\gamma =4/3-\epsilon$ with a
 small $\epsilon >0$ and discuss the physical
 implication of $\epsilon$.
We propose possible situations to sustain
 such void expansions in astrophysical contexts.

This paper is structured as follows. Section
 \ref{sec:INTRODUCTION} introduces background
 information, the physical problem, and our
 research motivation.
In Section \ref{sec:FORMULATION}, we show and discuss the
 formulation of self-similar transformation of hydrodynamic
 equations, as well as solution behaviours near sonic critical
 points and shock jump conditions.
Section \ref{sec:SOLUTIONS} discusses the behaviours of solutions
 near the void boundaries and gives various conventional polytropic
 self-similar void solutions crossing the sonic critical surface
 by different manners; shock solutions with different kinds of
 dynamic envelopes are presented.
We shall explore applications of those models especially
 in the context of SN in Section \ref{sec:APPLICATIONS}
 and some examples are presented there.
Section \ref{sec:CONCLUSION} contains
 conclusions and related discussions.
Appendices A to E offer details of physical
 reasoning and some mathematical derivations.

%{\bf Need to prepare an Introduction section to provide
%  background information and research motivation etc.
%  Explain the basic physical idea.}
%
%{\bf Please check the references of hot bubbles.
%  Emphasize what is different here!! }

\section{Formulation of Self-similar Model}
\label{sec:FORMULATION}

%{\bf Please make specific and sensible statements here.}

We first present below the nonlinear Euler hydrodynamic
  partial differential equations (PDEs) with spherical
  symmetry and self-gravity.
In spherical polar coordinates in which a spatial point is
  defined by $(r,\ \theta,\ \phi)$ where $r$ is the radius,
  $\theta$ is the polar angle and $\phi$ is the azimuthal
  angle, we will introduce self-similar transformation for
  a conventional polytropic gas
  \cite[e.g.][]{1988ApJ...326..527S, 2008MNRAS.390.1619H}.
For spherically symmetric hydrodynamics, all the dependent
  physical variables do not vary with angles $\theta$
  and $\phi$.

\subsection{Euler hydrodynamic PDEs,
  conventional polytropic self-similar transformation,\\
\qquad\  and asymptotic solutions}
  \label{sec:DYNAMIC_EQS_SIMILARITY_TRANS}

%{\it
With spherical symmetry and self-gravity,
  we have the radial momentum equation as
%, due to spherical symmetry: } {\bf Incorrect!}:
\begin{equation}
\label{eq:EULER_EQ} \dfrac{\partial u}{\partial t}
 + u\dfrac{\partial u}{\partial r} =
 -\dfrac{1}{\rho}\dfrac{\partial
 p}{\partial r} - \dfrac{G M }{r^2}\ ,
\end{equation}
where we use $u$ to denote the radial flow
 velocity, $t$ for the time, $p$ for
 the pressure, $\rho$ for the mass density,
%{\bf Define all notations!}
 $M$ for the enclosed mass within radius $r$ at time $t$,
 and $G=6.67\times 10^{-8}
 \mbox{ dyn }\mbox{cm}^2\mbox{ g}^{-2}$ for
 the universal gravitational constant.
%{\bf Specific value?}.
Nonlinear PDE \eqref{eq:EULER_EQ}
%, {\it a non-linear partial differential
%equation (PDE)} {\bf Define notations!},
is coupled with the mass conservation
\begin{equation}
\label{eq:MASS_CONSERVATION}
 \dfrac{\partial \rho}{\partial t}+
 \dfrac{1}{r^2}\dfrac{\partial}{\partial r}\left(\rho u r^2
 \right)=0\ ,
\end{equation}
or equivalently, in the integral form of
\begin{equation}
  \begin{split}
    &\dfrac{\partial M}{\partial t}+u\dfrac{\partial M}
    {\partial r}=0\ ,
    \\
    &\dfrac{\partial M}{\partial r}=4\pi r^2\rho\ .
  \end{split}
\end{equation}
%{\bf Spherical symmetry and more general hydrodynamic equations
%  are mixed? Check the consistency!}{\it Modified now.}
A closed set of nonlinear PDEs is established
 when we include the conventional polytropic EoS,
\begin{equation}
 \label{eq:POLYTROPIC_EOS}
    p=\kappa\rho^\gamma\ ,
\end{equation}
where $\gamma$ is a constant
 index and $\kappa$ is a global constant.
%{\bf Define notations when they are used for the first time!}

A self-similar transformation based on dimensional
 analysis [e.g. \citet{1988ApJ...326..527S}] for a
 conventional polytropic gas is introduced below, namely
\begin{equation}
\label{eq:SELF_SIMILAR_TRANSFORMATION}
\begin{split}
  & x=\dfrac{r}{k^{1/2}t^n}\ ,\quad u(r,t)=k^{1/2}t^{n-1}v(x)\
  ,\quad\rho (r,t)=\dfrac{\alpha (x)}{4\pi G t^2}\ ,
  \\
  & p(r,\ t)=\dfrac{k t^{2n-4}}{4\pi G}[\alpha (x)]^\gamma ,\quad
  M(r,\ t)=\dfrac{k^{3/2}t^{3n-2}}{(3n-2)G}m(x)\ ,
\end{split}
\end{equation}
where $k$ is the sound parameter and $n$ is the scaling exponent,
 $x$ is the dimensionless independent similarity variable,
 and depending on $x$ only, $v(x)$, $\alpha(x)$ and $m(x)$ are
 the dimensionless reduced variables for radial flow velocity
 $u(r,\ t)$, mass density $\rho(r,\ t)$ and enclosed mass
 $M(r,\ t)$, respectively.
Under transformation (\ref{eq:SELF_SIMILAR_TRANSFORMATION}),
 nonlinear PDEs \eqref{eq:EULER_EQ} to \eqref {eq:POLYTROPIC_EOS}
 %{eq:MASS_CONSERVATION}
%{\bf Please refer to them by numbers more specifically }
 are converted into two coupled
 nonlinear ordinary differential equations (ODEs), viz.,
\begin{equation}
\label{eq:DYNAMIC_ODES}
\begin{split}
  &\dfrac{\mbox{d}\alpha}{\mbox{d}x}=\dfrac{\alpha}{(nx-v)^2-
    \gamma\alpha^{\gamma -1}}
  \\
  &\ \times\left[(n-1)v + \dfrac{(nx-v)}{(3n-2)}\alpha
   -2\dfrac{(x-v)(nx-v)}{x}\right] ,
  \\
  &\dfrac{\mbox{d}v}{\mbox{d}x}=\dfrac{1}{(nx-v)^2-\gamma
    \alpha^{\gamma -1}}
  \\
  &\times\left[ (nx-v)(n-1)v + \dfrac{(nx-v)^2}{(3n-2)}\alpha
   -2\gamma\dfrac{x-v}{x}\alpha^{\gamma-1}\right] .
\end{split}
\end{equation}
According to \citet{1988ApJ...326..527S}, the conventional
  polytropic condition for a time-independent $\kappa$ in polytropic
  relation \eqref {eq:POLYTROPIC_EOS} simply requires $n+\gamma=2$.

With $n+\gamma=2$ and the assumption that $|v(x)|$ and $\alpha(x)$
  are non-increasing functions of $x$ at very large $x$,
  % ($x\gg 1$),
  $v(x)$ and $\alpha(x)$ have the following asymptotic
  behaviours\footnote{In
  \citet{1988ApJ...326..527S} the term
  proportional to $B^2$ has been missed. Lou \& Shi (2011
  in preparation) first pointed out this missing term
  in the asymptotic expansion for very large $x$.}
%However, this absence is not very important, since it is a term of
%  the order $x^{1-2/n}$, which is the second highest term in the
%  asymptotic solution.}
  in the regime $x\gg 1$, namely
\begin{equation}
\label{eq:ASYMPTOTIC_SOLUTIONS}
\begin{split}
  \alpha & = Ax^{-2/n}\ ,
  \\
  v & = A x^{(n-2)/n}\left[\dfrac{2(2-n)}{nA^n} -
    \dfrac{n}{(3n-2)}\right]
  \\
  & \qquad\qquad
  -\dfrac{(n-1)}{n}B^2x^{(n-2)/n}+B x^{(n-1)/n}
\end{split}
\end{equation}
%  {\bf One extra term missing in the second equation!!}
(Lou \& Shi 2011 in preparation),
%  \citep[][Lou \& Shi 2011 in preparation]{1988ApJ...326..527S}
  where $A$ and $B$ referred to as mass and speed parameters
  are two integration constants,
  characterizing asymptotic behaviours at large $x$.

The isothermal case of $n=1$ and $B=0$ was studied by
 \citet{1977ApJ...214..488S}, and a generalization of $B\neq 0$ for
 an isothermal gas was pursued by \citet{1985MNRAS.214....1W}
 for solutions with weak discontinuities (Lou \& Shen 2004).

Clearly, for all $n>0$ and $B\neq 0$, $Bx^{(n-1)/n}$
 is the leading term in the asymptotic solution of
 $v(x)$ for large $x$.
% {\bf Pay attention to the missing term proportional to $B^2$!}
Specific for isothermal cases (i.e. $n=1$ and $\gamma=1$),
 $B$ actually represents the constant reduced speed at
 $x\rightarrow +\infty$ \citep[e.g.][]{2004MNRAS.348..717L}.
In reference to our previous work
 \citep[e.g.][]{2006MNRAS.372..885L}, we shall refer
 to solutions according to their large$-x$ asymptotic
 envelope behaviours as follows in this paper:
% (note that they are discussed at large $x$).
$B=0$ and $v>0$ for ``breeze solutions"; $B=0$ and $v<0$ for
 ``contraction solutions"; $B>0$ (thus $v>0$ for large enough
 $x$) for ``outflow solutions"; $B<0$ (thus $v<0$ for large
 enough $x$) for ``inflow solutions".

Similarity transformation
 \eqref{eq:SELF_SIMILAR_TRANSFORMATION}
%{\it eqs. \eqref {eq:SELF_SIMILAR_TRANSFORMATION},}
%{\bf Provide a number here!  }
 and hydrodynamic PDEs lead to the following algebraic relation
\begin{equation}
\label{eq:MASS_CONDITION}
  m=\alpha x^2(nx-v)\ ,
\end{equation}
which implies the presence of the so-called ``zero-mass line"
 (ZML) specified by $nx-v=0$ in the $-v(x)$ versus $x$
 figure presentation; below this ZML a negative enclosed
 mass $M(r,\ t)$ would be physically unacceptable.

\subsection{The sonic critical curve and eigensolutions}
\label{sec:SCC}

% This is a newly inserted subsection. I
% am not using italic in this subsection.

In two coupled nonlinear hydrodynamic ODEs
 \eqref{eq:DYNAMIC_ODES}, we encounter a sonic critical point
 when the denominators on the right-hand side (RHS) vanish
 \citep[][n.b. $n+\gamma=2$]{1988ApJ...326..527S}, i.e.
\begin{equation}
\label{eq:SCC_DENOMINATOR}
  nx-v=(2-n)^{1/2}\alpha^{(1-n)/2}\ .
\end{equation}
In the three-dimensional (3D) variable space of $x$, $v$ and
 $\alpha$, condition (\ref{eq:SCC_DENOMINATOR}) represents
 a two-dimensional (2D) surface of singularity for two
 coupled nonlinear ODEs \eqref {eq:DYNAMIC_ODES}.
In general, a solution cannot go across this surface smoothly
 due to the singularity of ODEs except for a special
 type of solutions, viz., the eigensolutions.
One important necessary condition for searching such
 eigensolutions is that the numerators on the RHS
 also vanish simultaneously, which is equivalent to
 \citep[see also][]{1988ApJ...326..527S}
\begin{equation}
\label{eq:SCC_NUMERATOR}
 (n-1)v+\dfrac{(nx-v)}{(3n-2)}\alpha
 -2\dfrac{(x-v)(nx-v)}{x}=0\ .
\end{equation}
While requirement \eqref{eq:SCC_NUMERATOR} is derived to have
 the numerator of $\mbox{d}\alpha / \mbox{d} x$ to vanish,
 it is straightforward to verify that the numerator of
 $\mbox{d} v/\mbox{d}x$ in eq. \eqref{eq:DYNAMIC_ODES}
 also vanishes with eq. \eqref{eq:SCC_DENOMINATOR} given
 \citep[see][Lou \& Shi 2011 in preparation]{1988ApJ...326..527S}.
In other words, conditions \eqref{eq:SCC_DENOMINATOR} and
 \eqref{eq:SCC_NUMERATOR} are sufficient to determine those
 eigensolutions crossing the sonic critical curve (SCC)
 smoothly, leaving one degree of freedom (DOF), i.e. the
 specific point at which the eigensolutions go across.
% {\bf How about the second one in (6)?
% Please demonstrate the equivalence.}
ODEs \eqref{eq:SCC_DENOMINATOR} and
 \eqref{eq:SCC_NUMERATOR}
 are combined to determine the SCC in the 3D
 variable space of $x$, $v$ and $\alpha$.
The points at which the eigensolutions cross the critical
 surface smoothly are all located along the SCC.
In the following, we shall use $x_0$, $v_0$ and $\alpha_0$
 to denote values of $x$, $v$ and $\alpha$ on the SCC for
 solutions crossing the SCC smoothly.
When solving two ODEs \eqref{eq:SCC_DENOMINATOR} and
 \eqref{eq:SCC_NUMERATOR} explicitly for $x_0$, $v_0$ and
 $\alpha_0$, we find that the SCC consists of two segments
 represented by subscripts $+$ and $-$ respectively as two
 connected quadratic solutions:
\begin{equation}
\label{eq:SCC_SOLUTIONS}
  \begin{split}
    x_{0,\pm}&=\dfrac{-K_b\pm (K_b^2-4K_aK_c)^{1/2}}{2K_a}\ ,
    \\
    v_{0,\pm}&=n x_{0,\pm}-(2-n)^{1/2}\alpha_0^{(1-n)/2}\ ,
  \end{split}
\end{equation}
where
\begin{equation}
  \begin{split}
    K_a &\equiv n(n-1)\ ,
    \\
    K_b &\equiv (2-n)^{1/2}\left(n-1+\dfrac{\alpha_0}{3n-2}\right)
    \alpha_0^{(1-n)/2}\ ,
    \\
    K_c &\equiv 2(n-2)\alpha_0^{1-n}
  \end{split}
\end{equation}
\citep{2008MNRAS.384..611L}. These two $+$ and $-$
  segments of the SCC meet right at the point where
\begin{equation}
  \alpha_0=-3n^2+5n-2+2\big[2(4n-16n^2+21n^3-9n^4)\big]^{1/2}.
\end{equation}
We combine these two SCC branches to form a smooth global SCC.
 In the following, we refer to the segment with ``$+$''
  in eq. \eqref{eq:SCC_SOLUTIONS} as segment 1, while that
  with ``$-$'' in eq. \eqref{eq:SCC_SOLUTIONS} as segment 2,
  respectively.

We then expand an eigensolution to the first order near the SCC as
\begin{equation}
  x = x_0+\delta\ ,\qquad\alpha=\alpha_0+\alpha'\delta\ ,
  \qquad v =v_0 + v'\delta\ ,
\end{equation}
where $\alpha'$ and $v'$ are the values of $\mbox{d}
 \alpha/\mbox{d}x$ and $\mbox{d}v/\mbox{d}x$ on the
 SCC for eigensolutions, respectively.
Solving two ODEs (\ref{eq:SCC_DENOMINATOR}) and
 (\ref{eq:SCC_NUMERATOR}), we obtain a quadratic
 equation of $v'$ as well as the equation of $\alpha'$
 \citep[see also][]{1988ApJ...326..527S}:
\begin{equation}
\label{eq:SCC_DERIVATIVE}a v'^2 + b v' + d = 0\ ,\quad
 \alpha' =\dfrac{\alpha_0^{(n+1)/2}}
 {(2-n)^{1/2}}\left( v' - 2\dfrac{x_0-v_0}{x_0}\right),
\end{equation}
where three coefficients $a$, $b$ and $c$
  are explicitly defined by
\begin{equation}
\begin{split}
  a & \equiv 3 - n \ ,
  \\
  b & \equiv 4(1-n)\dfrac{v_0}{x_0}+5n-7\ ,
  \\
  d & \equiv 2(3-2n)\dfrac{v_0^2}{x_0^2}+
    2\left(\dfrac{\alpha_0}{3n-2}+
    2n-4\right)\dfrac{v_0}{x_0}
  \\
  &\qquad\qquad\qquad\qquad +\dfrac{(n-2)}{(3n-2)}\alpha_0 -2n+ 4\ .
\end{split}
\end{equation}
Quadratic equation (\ref{eq:SCC_DERIVATIVE}) for $v'$
 generally allows two roots, corresponding to two types
 of possible eigensolutions across the SCC at the same
 point on the SCC \citep{2008MNRAS.384..611L}.
To be specific, with the formula of roots of
 quadratic equation (\ref{eq:SCC_DERIVATIVE}),
 we obtain the two roots of $v'$ explicitly as
\begin{equation}
  \label{eq:V_PRIME}
  v'_\pm =\dfrac{-b\pm (b^2-4ad)^{1/2}}{2a}\ .
\end{equation}
For an eigensolution crossing the SCC only once, those with
 $v'_-$ will be referred to as type-1 solutions while those
 with $v'_+$ are type-2 solutions.
In general, those sonic critical points at which the solution
 take $v'=v'_-$ will be referred to as type-1 critical point,
 while $v'=v'_+$ for type-2 critical point.

\subsection{Shock jump conditions across the SCC}
\label{sec:SHOCK}

For astrophysical applications, it is common to expect
 the existence of shocks in the hydrodynamic evolution
 of a gas sphere.
Across a shock front propagating in a conventional
 polytropic gas, conditions of the mass conservation
\begin{equation}
\label{eq:SHOCK_MASS_CONSERVATION}
  (\rho_2-\rho_1)u_s=u_2\rho_2-u_1\rho _1\ ,
\end{equation}
the radial momentum conservation
\begin{equation}
\label{eq:SHOCK_MOMENTUM_CONSERVATION}
  u_s(u_2 \rho _2-u_1 \rho _1)
  = -p_1+p_2-u_1^2\rho _1+u_2^2 \rho _2\ ,
\end{equation}
and the energy conservation
% {\bf the first two terms on the LHS of eq (11)
% below missed factor $\gamma$?}
\begin{equation}
\label{eq:SHOCK_ENERGY_CONSERVATION}
\begin{split}
  & u_s\left(-\dfrac{p_1}{\gamma -1}+
    \dfrac{p_2}{\gamma-1}-\frac{u_1^2 \rho _1}{2}
    +\frac{u_2^2 \rho_2}{2}\right)
  \\
  & = u_2\left(\dfrac{\gamma p_2 }{\gamma -1}+\frac{u_2^2\rho
      _2}{2}\right)-u_1\left(\dfrac{\gamma
      p_1}{\gamma-1}+\frac{u_1^2\rho_1}{2}\right)\ ,
\end{split}
\end{equation}
need to be satisfied in the co-moving shock reference
 framework in order to attain a physically acceptable
 shock discontinuity.
We note here, especially for the energy conservation across
  the shock front, that the polytropic index $\gamma$ should
  remain the same across a shock front for similarity solutions;
if those $\gamma$ are different, $n$
  would be also different because of $n+\gamma=2$.
This would not be acceptable for a global similarity
  evolution in which a shock evolves self-similarly.
%{\bf In general, we could also allow $\gamma$ to be
%different across a shock for a general polytropic gas!}

It is useful to point out that those three shock equations
  above are invariant under a commutation of subscripts 1
  and 2.
We need to know the direction for entropy increase
  for identifying physical shock solutions.
To be specific, we denote subscript 1 for the upstream flow.
Solving combined shock conditions
  \eqref{eq:SHOCK_MASS_CONSERVATION}$-$\eqref{eq:SHOCK_ENERGY_CONSERVATION}
  and adopting self-similar transformation
  \eqref{eq:SELF_SIMILAR_TRANSFORMATION}, those conservation
  conditions become explicitly ($x_{\text{s1}}$ is the
  self-similar location of the shock on the upstream side
  and $x_{\text{s2}}$ being that of the downstream side;
  in fact, they correspond to the same shock radius $r_s(t)$
  but with different $k$ values for the sound parameter)
\begin{equation}
\label{eq:SHOCK_SOLUTIONS}
\iffalse
    x_\text{s2} =& x_\text{s1} \Bigg\lbrace \frac{2\alpha _1
    \left[v_1-n x_\text{s1}\right)^2-(\gamma -1)
    \alpha_1^\gamma}{\gamma +1}
    \\
    \times &\left[\frac{\alpha _1^2 (\gamma +1)
    \left(v_1-n x_{\text{s1}}\right)^2}{\alpha_1(\gamma -1)
    \left(v_1-n x_{\text{s1}}\right)^2+2\gamma
    \alpha_1^{\gamma }}\right]^{-\gamma }\Bigg\rbrace^{-1/2}\ ,
    \\
    v_2 = & \Bigg\lbrace \dfrac{\left(v_1-n x_\text{s1}\right)^2}
    {\alpha_1^2(\gamma +1)\left(v_1-n x_\text{s1}\right)^2}
    \dfrac{\left[\alpha _1\left(2 n x_\text{s1}+v_1\gamma
    -v_1\right)+2 \gamma  \alpha _1^\gamma\right]^2}{2\alpha_1
    \left(v_1 - nx_\text{s1}\right)^2-(\gamma -1)\alpha_1^\gamma}
    \\
    \times & \left[\frac{\alpha_1^2 (\gamma +1)
    \left(v_1-n x_{\text{s1}}\right)^2}{\alpha_1 (\gamma -1)
    \left(v_1-n x_{\text{s1}}\right)^2
    +2\gamma\alpha_1^{\gamma }}\right]^\gamma\Bigg\rbrace^{1/2}\ ,
    \\
\fi
\begin{split}
  & \alpha _2 =\frac{\alpha_1^2 (\gamma +1) \left(v_1-n
      x_{\text{s1}}\right)^2}{\alpha _1 (\gamma -1)\left(v_1-n
      x_{\text{s1}}\right)^2+2\gamma \alpha _1^\gamma }\ ,
  \\
  & x_{s2} = x_{s1}\bigg[
    \dfrac{\alpha_2^\gamma(\gamma+1)}{2(v-nx_{s1})^2 -\alpha_1^\gamma
      (\gamma-1)} \bigg]^{1/2}\ ,
  \\
  & v_2 = \dfrac{ 2nx_{s1} + v_1 (\gamma-1) + \dfrac{ 2\gamma \alpha
      _1 ^{\gamma-1} }{(v_1-x_{s1})} }{(\gamma+1)}
      \dfrac{x_{s2}}{x_{s1}}\ ,
\end{split}
\end{equation}
which are sufficient to determine those important
 self-similar variables \{$x$, $v$, $\alpha$\} at the
 immediate downstream side of a shock front when those
 of the immediate upstream side are known.
The other way around, it is also straightforward to
 determine physical variables of the upstream side when
 physical variables of the downstream side are specified.

In addition, we introduce parameter $\eta_k$ for convenience
 as the ratio of different sound parameter $k$ values at the
 upstream to downstream sides across a shock front, namely
\begin{equation}
  \eta_k =\dfrac{k_1}{k_2}=
  \left(\dfrac{x_{s2}}{x_{s1}}\right)^2
  =\dfrac{\alpha_2^\gamma(\gamma+1)}
  {2(v-nx_{s1})^2 -\alpha_1^\gamma (\gamma-1)}\ .
\end{equation}
Apparently, $\eta_k$ reflects the strength of an
 expanding shock in a self-similar dynamic evolution.

%{\it
This $\eta_k$ parameter here is equivalent to $1/\lambda$
 in the formulation of \citet {2006MNRAS.372..885L}.
We now discuss the variation of specific entropy along
 streamlines, reflected by the variation of $k$ (or $\kappa$;
 $\kappa=k(4\pi G)^{\gamma-1}t^{2(n+\gamma-2)}$ for a
 general polytropic gas and $\kappa=k(4\pi G)^{1-n}$ for a
 conventional polytropic gas) across an outgoing shock front.
The Mach number $\mathcal{M}_i$ of a shock is defined by
\begin{equation}
  \mathcal{M}_i=|u_s-u_i|/s_i\ ,
\end{equation}
where $i=1,\ 2$ refer to immediate upstream and downstream sides,
%and in the following context denotes the
%upstream and downstream side of the shock
 respectively, and $u_i$ is the radial flow speed
 near the shock front, $s_i$ is the sound speed,
 $u_s$ is the expansion speed of the shock front
 in our frame of reference.
Using similarity transformation
 \eqref{eq:SELF_SIMILAR_TRANSFORMATION},
 we derive the Mach number of a shock in self-similar
 form on either side of a shock front as
\begin{equation}
  \mathcal{M}_i =\dfrac{x_i^2}{\gamma
  \alpha_i^{\gamma-1}}\Gamma_i\ ,
\end{equation}
where notations $\Gamma_i$ and $z$ are defined by
\begin{equation}
  \Gamma_i\equiv n -v_i/x_i\ ,
  \qquad\quad z\equiv\Gamma_2/\Gamma_1\ ,
\end{equation}
which are the same as those in
 \citet{2006MNRAS.372..885L}.\footnote{There is a typo in
 the definition of $\Gamma_i$ in \citet{2006MNRAS.372..885L}
 which is now corrected here.}
In the following discussion, we assume $\gamma>1$.
 It is then straightforward to show that when
 $\mathcal{M}_1>1$, we must have $z<1$.
Since $z<1$ gives (see also \citealt{2006MNRAS.372..885L})
\begin{equation}
  \dfrac{\mbox{d}(\lambda^2)}{\mbox{d}z}
  =\dfrac{\mbox{d}}{\mbox{d}z}\bigg(\dfrac{1}{\eta_k^2}\bigg)
  = - \dfrac{\gamma(\gamma^2-1)(z-1)^2
    z^{\gamma-1}}{\left[(\gamma+1)z-(\gamma-1)\right]^2} < 0\ ,
\end{equation}
and $\lambda\rightarrow 1$ when $z\rightarrow 1$, this derivation
 yields that when $\mathcal{M}_1>1$, inequalities $\lambda>1$ and
 $\eta_k<1$ (thus $k_1<k_2$ and $\kappa_1<\kappa_2$) will hold.
This indicates the fact that the shock wave is supersonic in the
 upstream flow ($\mathcal{M}_1>1$) and causes an increase of
 entropy across a shock front (i.e., $\kappa_2>\kappa_1$).
If however $\mathcal{M}_1$ is less than $1$, subscript $1$ would
 therefore denote downstream and the condition of entropy increase
 still holds (i.e., $\kappa_2<\kappa_1$).
In summary, shock jump conditions
 \eqref{eq:SHOCK_MASS_CONSERVATION}$-$\eqref
 {eq:SHOCK_ENERGY_CONSERVATION} and naturally their solutions
 \eqref{eq:SHOCK_SOLUTIONS} are invariant with respect to a
 commutation of subscripts $1$ and $2$; the irreversibility
 of a shock or the increase of specific entropy across a
 shock front should not be violated.
% }

\section{Conventional Polytropic Self-similar Void Solutions }
\label{sec:SOLUTIONS}

\begin{figure}
  \includegraphics[bb=0 0 330 265, width=70mm]{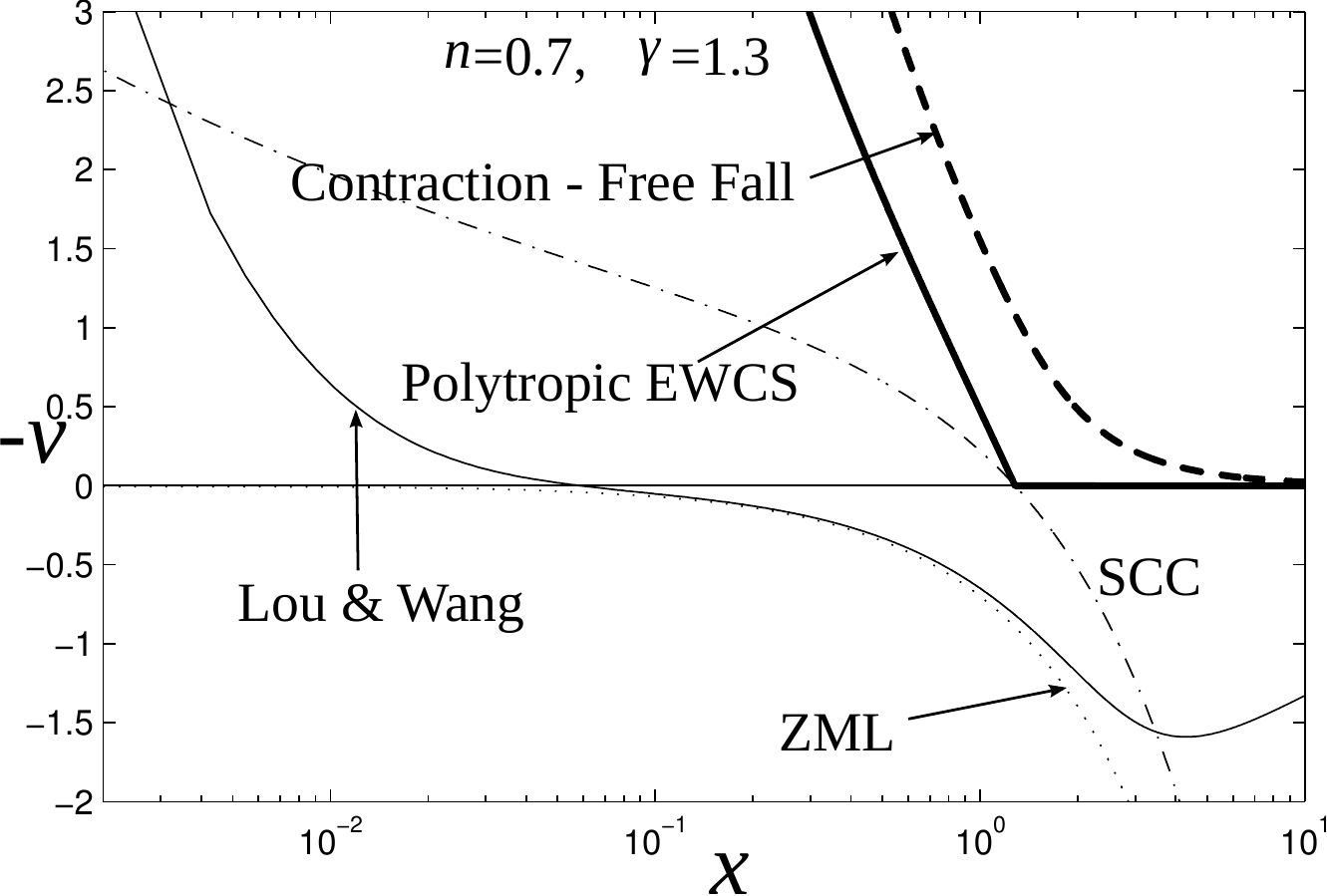}
  \caption{
  Examples of conventional polytropic self-similar
     solutions \citep[][]{2006MNRAS.372..885L}.
%    \citep[reproduced from][]{2006MNRAS.372..885L}.
  These solutions have $n=0.7$ (thus $\gamma=1.3$).
  For a clearer illustration, the horizontal $x$-axis
    is shown in a logarithmic scale.
  The dash-dotted curve is the SCC, shown as
    a reference here.
%, which has been well-defined only for the solutions
%    going across it smoothly.
  The light solid curve marked by ``Lou \& Wang" is the solution that
    crosses the SCC twice.
  It has two critical points on the SCC at $(\alpha,\ x,\ v)=(164.69,\
    0.002978,\ -2.4497)$ for a type-2 crossing and $(\alpha,\ x,\ v)=
    (0.1639,\ 3.4684,\ 1.5586)$ for a type-1 crossing, respectively
    (see \citet{2006MNRAS.372..885L} for more details).
  The heavy dashed curve marked by ``Contraction - Free Fall'' is the
    solution with contraction envelope and a free-fall collapsing core
    with asymptotic behaviour $A=A_\text{e}+0.3$ ($A_\text{e}$ is
    given by eq. \eqref{eq:A_E}), a polytropic counterpart of
    similar isothermal results in \citet{1977ApJ...214..488S}.
  The heavy solid curve is the polytropic EWCS solution
    ($A\rightarrow A_\text{e}^+$).
  The dotted curve is the zero-mass line (ZML with $nx=v$).
%  {\bf Not sufficiently clear for marked names. }
  }\label{fig:SHU_LS_0_7}
\end{figure}

\subsection{Several pertinent self-similar solutions
  presented in \citet{2006MNRAS.372..885L} and\\
  \qquad\ \citet{2008MNRAS.384..611L}}

We first illustrate here several related self-similar
 solutions for conventional polytropic gas flows
 already obtained earlier.
These are important in two aspects, viz., validating
 our numerical code and preparing for further model
 analysis for self-similar voids surrounded by
 various dynamic envelopes.

\subsubsection{Conventional polytropic expansion-wave\\
\qquad\quad  collapse solution (EWCS)}

Expansion-wave collapse solutions (EWCSs) were studied originally
 in an isothermal gas by \citet{1977ApJ...214..488S}.
The natural generalization to conventional polytropic
 EoS was done by \citet{1978ApJ...221..320C}.
With a general polytropic EoS, this kind of solutions with
 $\gamma=4/3$ and $n+\gamma\neq 2$ has been constructed
 by \citet{2008MNRAS.384..611L} in recent years.

Conditions for EWCS can be readily derived from
 self-similar hydrodynamic ODE \eqref{eq:DYNAMIC_ODES};
 meanwhile, those for the outer portion of a static
 envelope of a singular polytropic sphere (SPS) have
 been constructed \citep[e.g.][]{2006MNRAS.372..885L}.
We can independently use some practical methods to obtain
 these solutions numerically with asymptotic solutions
 \eqref{eq:ASYMPTOTIC_SOLUTIONS} in the regime of large $x$.
%  {\bf Extra term!!} {\it The extra term will not
%  affect the discussion here in this subsection.}
As the coefficient of the leading term in the asymptotic
 expression of $v(x)$, speed parameter $B$ must vanish
 for this type of solution for a static polytropic envelope.
For mass parameter $A$, we can get it from the asymptotic
 solution of $v(x)$ by making the coefficient of the term
 of order $x^{(n-2)/n}$ to vanish, i.e.
\begin{equation}
  \left[\dfrac{2(2-n)}{nA^n} -
  \dfrac{n}{(3n-2)}\right]\rightarrow 0^-\ ,
\end{equation}
which then yields ($A_\text{e}$ is the limit for mass
 parameter $A$ to achieve the conventional polytropic EWCS)
\begin{equation}
\label{eq:A_E} A\rightarrow A_\text{e}^+\ ,
\qquad\qquad
A_\text{e} =\left[\dfrac{2(2-n)(3n-2)}{n^2}\right]^{1/n}\ .
\end{equation}
For the isothermal EWCS of Shu (1977), we
 have $n=1$ and $A_\text{e}=2$.
This $A_\text{e}$ value of $A$ represents a limit.
 An exact equality of $A=A_\text{e}$ would correspond to
 the SPS solution.
% and render numerical methods invalid due to numerical
% singularity. {\bf Please explain more explicitly!}

EWCS can be regarded as a special case bounding
 ``collapse solutions without critical points"
  with $A>A_\text{e}$
% {\bf touching the critical point}
  \citep[e.g.][]{1977ApJ...214..488S,
  springerlink:10.1007/s10509-009-0044-4}.
Here we present one of such solutions with $n=0.7$ in Fig.
 \ref{fig:SHU_LS_0_7} (the conventional polytropic relation
 $n+\gamma=2$ holds).
We found that the behaviours of solutions are qualitatively
 similar to isothermal cases as $A$ gradually goes down to
 $A_\text{e}=2$ \cite[see][]{2008MNRAS.384..611L}; here the
 limit of $A$ becomes $A_\text{e}= 0.4044$ for $n=0.7$.
For an exact conventional polytropic EWCS, a
 discontinuity in the first derivative appears at $x=1.3085$,
 where the central infall side solution is tangent to the SCC.
This discontinuity in first derivatives denotes the front of
 an expansion wave in self-similar evolution inside of which
 a collapse takes place and beyond which the gas remains a
 static conventional polytropic envelope (i.e. outer part
 of a SPS).
We shall further construct this kind of solutions
 in the following.

\subsubsection{Solutions crossing the SCC smoothly}

\begin{figure}
  \includegraphics[bb=0 0 330 505, width=70mm]
  {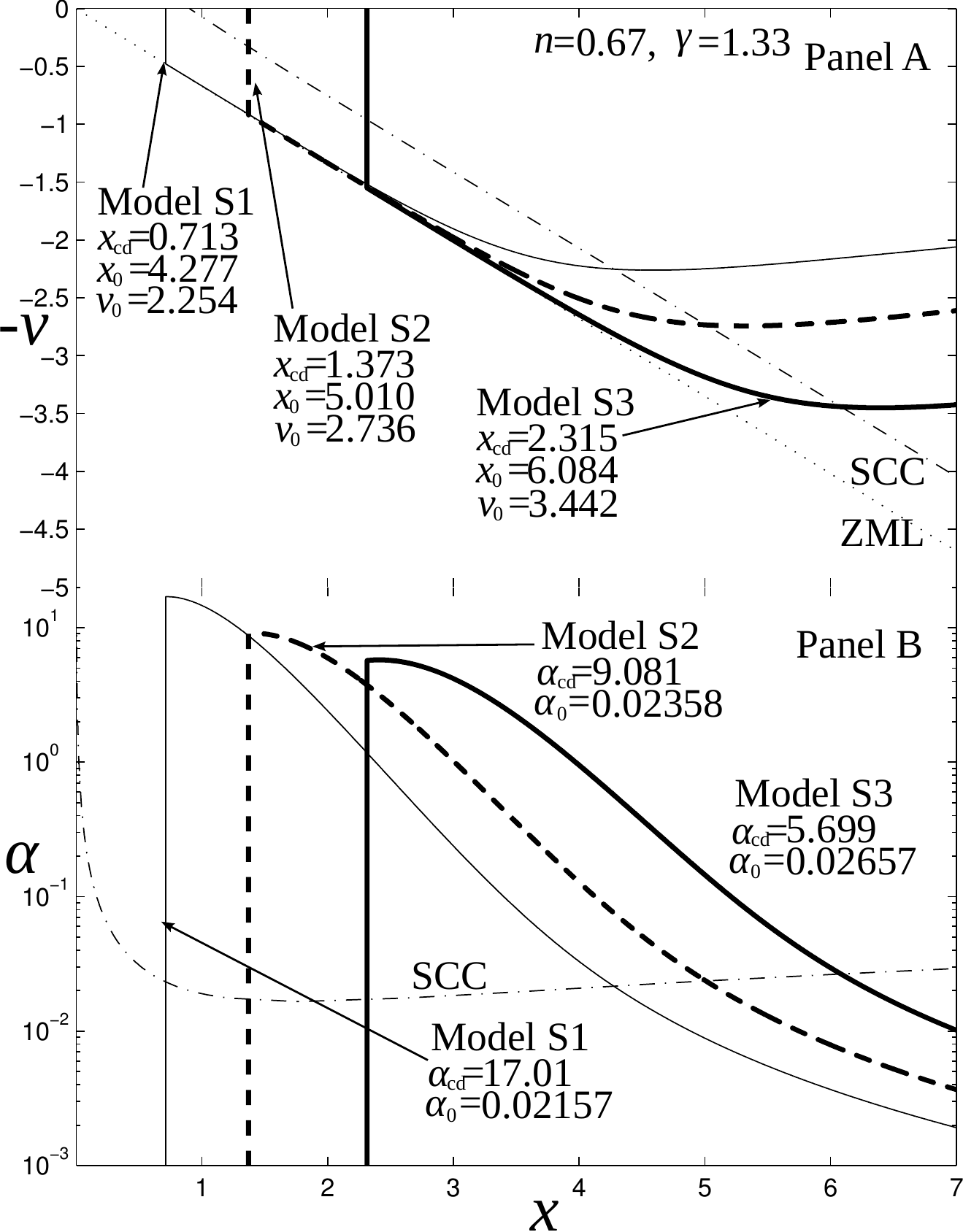}
  \caption{
% {\bf Modify SCC to SCC in the figures!! label $v_0$.}
Conventional polytropic void solutions crossing
    the SCC smoothly with $n=0.67$ (thus $\gamma=1.33$).
Panel A illustrates the profiles of negative reduced
    radial speed $-v(x)$ versus $x$ and Panel B shows
    the corresponding profiles of reduced mass density
    $\alpha(x)$ versus $x$ where the $\alpha$-axis is
    in the logarithmic scale.
The dash-dotted curves in both panels are the SCCs,
    while the light dotted line in Panel A is the ZML.
Three solutions of such type, marked by Model S1 to S3 (S
    for ``smooth'') are plotted in light solid curve, heavy
    dashed curve and heavy solid curve, respectively.
Model S1 crosses the SCC at $(x_0,\ v_0,\ \alpha_0) =
    (4.277,\ 2.254,\ 0.02157)$,
    Model S2 at $(x_0,\ v_0,\ \alpha_0)=(5.010,\ 2.736,\ 0.02358)$
    and Model S3 at $(x_0,\ v_0,\ \alpha_0)=(6.084,\ 3.442,\ 0.02657)$.
Specific data and parameters for each solution
    are labelled in the figure. }
    \label{fig:SCC_T1_0_67}
\end{figure}

We present here several eigensolutions crossing the SCC smoothly.
In our Fig. \ref{fig:SHU_LS_0_7}, the solution curve marked with
  ``Lou \& Wang" is such an eigensolution example taken from
  figure 1 of \citet{2006MNRAS.372..885L}.
This solution of $n=0.7$ (thus $\gamma =1.3$) is the
  conventional polytropic counterpart of isothermal
  envelope expansion core collapse (EECC) solutions
  of \citet{2004ApJ...611L.117S}.
Specific parameters are summarized in the
  caption of Fig. \ref{fig:SHU_LS_0_7}.

\subsection{Dynamic void solutions and
  their\\
  \qquad\ behaviours near the void boundary}

\begin{figure}
  \includegraphics[bb=0 0 330 280, width=70mm]
  {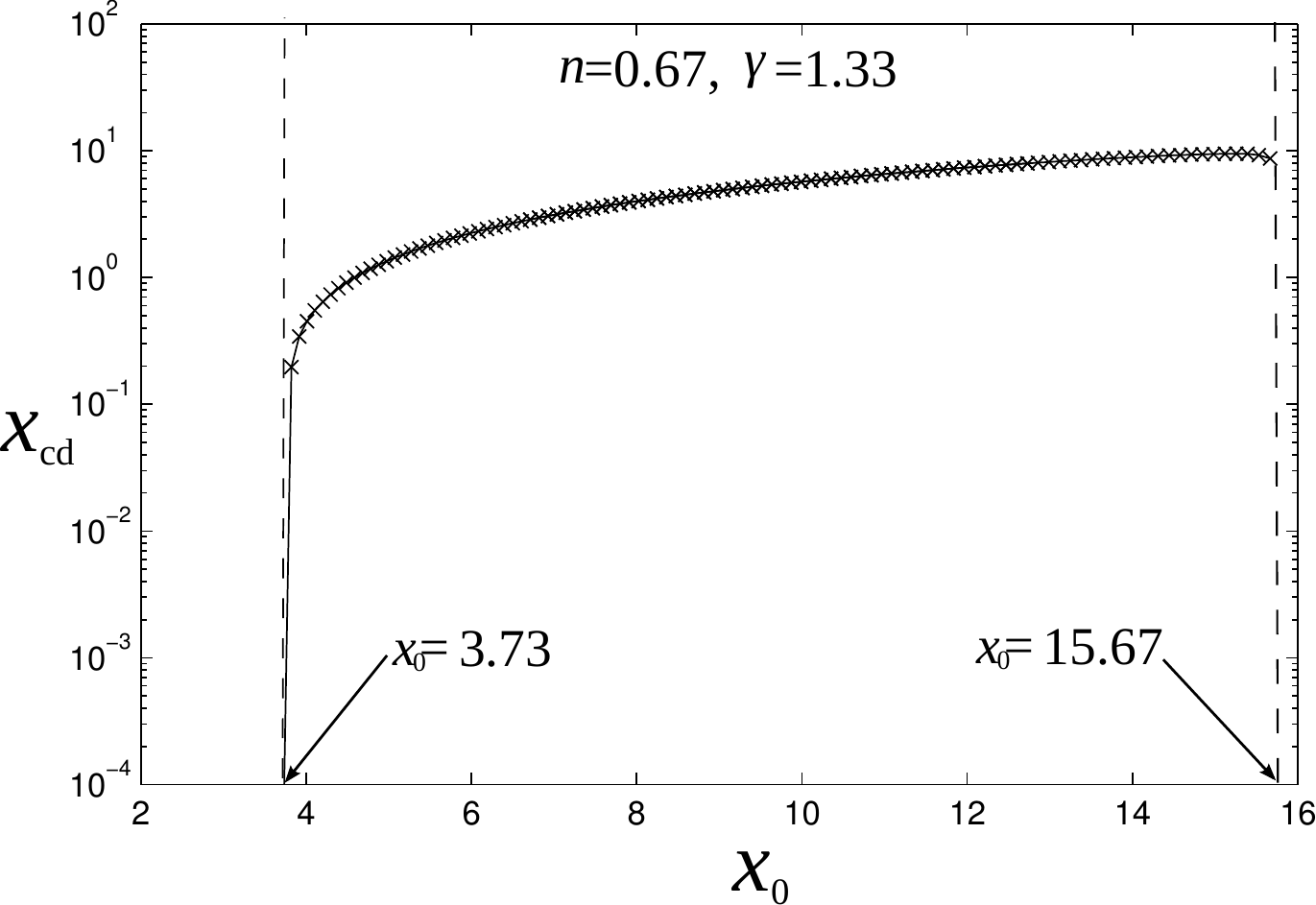}
  \caption{
%  {\bf Need proper labels!}
The relation of $x_\text{cd}$ (where $\alpha_\text{cd}$ and
  $v_\text{cd}$ can be determined) versus
%   $\alpha_\text{cd}$ and $v_\text{cd}$ for different
    $x_0$ for the conventional polytropic case of
    $n=0.67$ and $\gamma=1.33$.
The light solid curve marked along with crosses (``$\times$'')
    corresponds to a branch of type-1 solutions, which is the
    only possible solution branch for $x_0\lsim 50$. That is,
    there are no type-2 solutions to reach $x_\text{cd}$
    for $x_0\lsim 50$.
 %   {\bf Please explain more clearly.}
 }
    \label{fig:SCC_PD_0_67}
\end{figure}

\begin{figure}
  \includegraphics[bb=0 0 330 535, width=70mm]
  {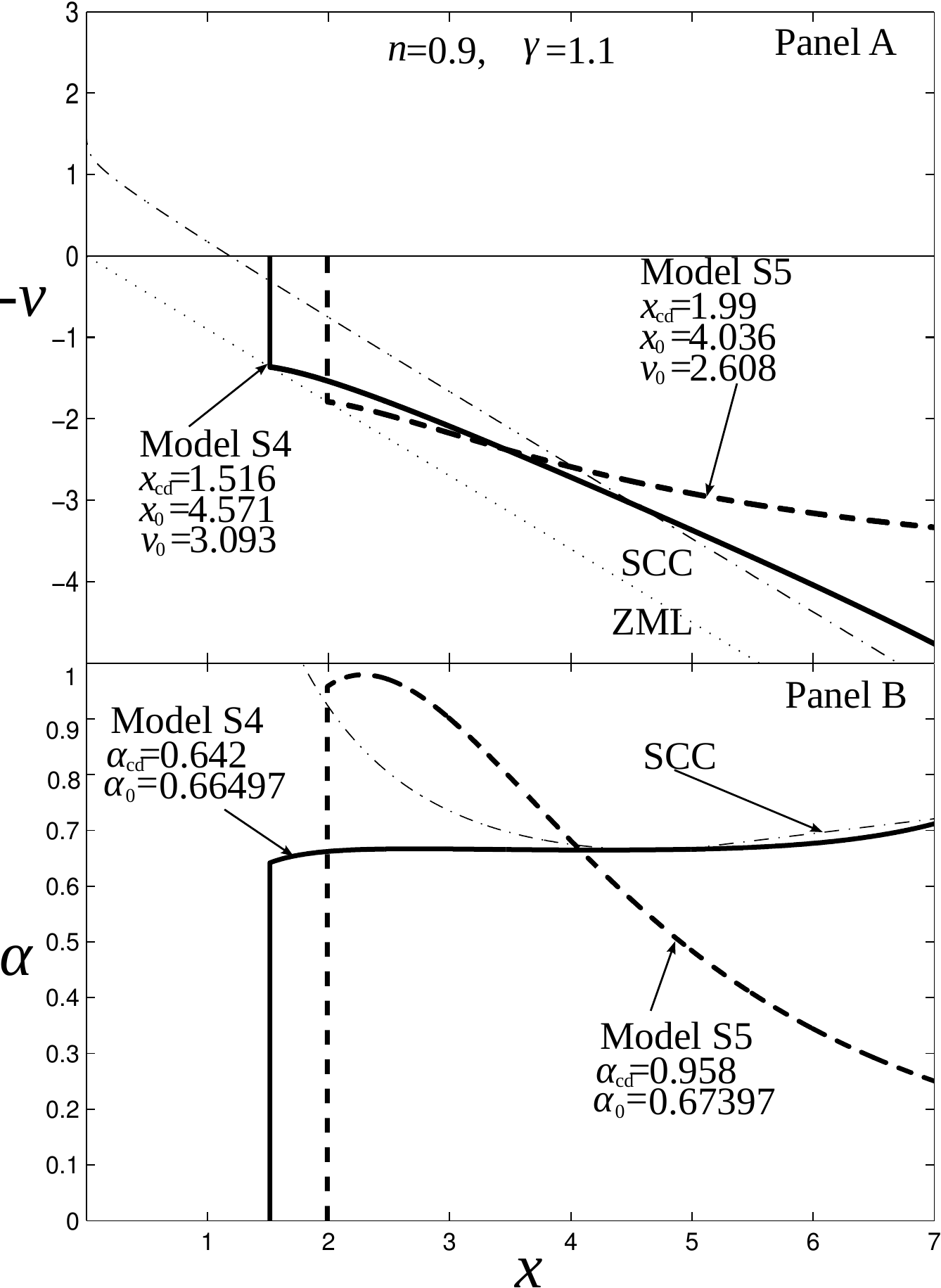}
  \caption{
Conventional polytropic void solutions crossing
  the SCC smoothly with $n=0.9$ ($\gamma=1.1$).
Panel A illustrates the profiles of negative reduced
  radial speed $-v(x)$ versus $x$ and Panel B shows
  the corresponding profiles of reduced mass density
  $\alpha(x)$ versus $x$.
The dash-dotted curves in both panels are the SCC,
  while the light dotted line in Panel A represents the ZML.
Two solutions of such type, marked by Model S4 and S5 (S for
  ``smooth") are plotted in heavy solid curve and heavy dashed
  curve, respectively.
Model S4 is a type-2 solution crossing the SCC at $(x_0,\ v_0,\
  \alpha_0)=(4.571,\ 3.093,\ 0.66497)$ and Model S5 is a type-1
  solution crossing the SCC at $(x_0,\ v_0,\ \alpha_0)=(4.036,\ 2.608,\
  0.67397)$.}\label{fig:SCC_0_9}
\end{figure}

Among our self-similar solutions, it is possible
 for some to reach a specific line
% in the $v$ versus $x$ profiles -- this line corresponds to
 $nx=v$ which is referred to as the ZML. By eq.
\eqref{eq:MASS_CONDITION}, the ZML leads to
\begin{equation}
  m|_{nx=v}=\alpha x^2(nx-v)=0\ ,
\end{equation}
and thus, according to eq.
 \eqref{eq:SELF_SIMILAR_TRANSFORMATION},
\begin{equation}
  M|_{nx=v} =\dfrac{k^{3/2}t^{3n-2}}{(3n-2)G}m|_{nx=v}=0\ .
\end{equation}
In other words, if there is a point in a solution where $v$ equals
 to $nx$, the enclosed mass $M$ within the corresponding radius
 of $nx=v$ (this radius expands with time $t$) would vanish.

On one hand, these would suggest that there is an expanding void,
 inside which mass and gravity could be neglected as compared
 with those of the surrounding gas shell, embedded inside
 a time-dependent and self-similarly evolving interface
 characterized by $nx=v$.
On the other hand, condition $nx=v$ implies that the
 expanding interface represents a contact discontinuity
 (e.g. Lou \& Zhai 2009, 2010) as revealed by the
 following relation
\begin{equation}
  \dfrac{\mbox{d}r}{\mbox{d}t}=nk^{1/2}x t^{n-1}
  =k^{1/2}v t^{n-1}=u\ ,
\end{equation}
at the time-dependent interface radius where $nx=v$.
 This is one of the necessary conditions for the existence
 of contact discontinuity \citep[e.g.][]{Lou2010198} and
 the other requirement is the balance of pressures on
 both sides, which will be discussed in
 Section \ref{sec:APPLICATIONS} presently.
In the following, we invoke subscript ``cd" to refer
 to those variables at the immediate outside of the contact
 discontinuity (for independent self-similar variable $x$
 as an example, $x_\text{cd}$ is simply the similarity
 location of the contact discontinuity surface).
For instance,
\begin{equation}
  \alpha_\text{cd}=\lim_{x\rightarrow
  x_\text{cd}^+}\alpha(x)\ .
\end{equation}
%
%It is necessary to have some discussion on
We note here several properties of solution
 adjacent to the void boundary from the
 gaseous envelope side.
According to self-similar hydrodynamic ODE
 \eqref{eq:DYNAMIC_ODES}, we can determine the first
 derivatives at the ZML, viz.
\begin{equation}
  \left. \dfrac{\mbox{d}v}{\mbox{d}x}\right|_\text{cd}=2(1-n)\ ,
  \qquad\quad
  \left. \dfrac{\mbox{d}\alpha}{\mbox{d}x} \right|_\text{cd}
  =\dfrac{n(1-n)}{(2-n)}\alpha_\text{cd}^n x_\text{cd}\ ,
\end{equation}
and therefore the Taylor series expansions near the ZML are
 \citep[this is actually the case of $q=0$
  for a conventional polytropic gas in][]{2008MNRAS.390.1619H,
  Lou2010198}
\begin{equation}
\begin{split}
  v(x) &=nx_\text{cd}+2(1-n)(x-x_\text{cd})+\cdots\ ,
  \\
  \alpha(x) &=\alpha_\text{cd}+\dfrac{n(1-n)}{(2-n)}
  \alpha_\text{cd}^n x_\text{cd}(x-x_\text{cd})+\cdots\ .
\end{split}
\end{equation}
%{\bf Please clarify the first expression!}
Apparently, $\mbox{d}\alpha/\mbox{d}x$ at the contact
 discontinuity vanishes if $\alpha_\text{cd}=0$.
This in turn hints that $\alpha(x)$ might be zero everywhere,
 and we have indeed verified this by extensive numerical
 explorations \citep[see also][]{Lou2010198}.
Moreover, $\alpha_\text{cd}=0$ leads to a singularity
 for the second order derivative of $v(x)$ there (see Appendix A).
In short, cases with $\alpha_\text{cd}=0$ are not physically
 acceptable for a conventional polytropic gas.
Meanwhile for $m_\text{cd}=0$, we must have
\begin{equation}
  \lim_{x\rightarrow x_\text{cd}^-}\alpha = 0\ ,
\end{equation}
indicating the presence of a jump or ``cliff" in mass
 density across a contact discontinuity (this is exactly
 what the term ``contact discontinuity" implies).
The difference in mass densities inevitably leads to
 diffusion of gas particles across the contact
 discontinuity surface, which we shall discuss
 in Appendix \ref{sec:DIFFUSION}.
We shall focus on void solution cases
 of non-vanishing $\alpha_\text{cd}>0$.

\subsection{Conventional polytropic
 void solutions crossing the SCC smoothly}

\begin{figure}
  \includegraphics[bb=0 0 330 280, width=70mm]
  {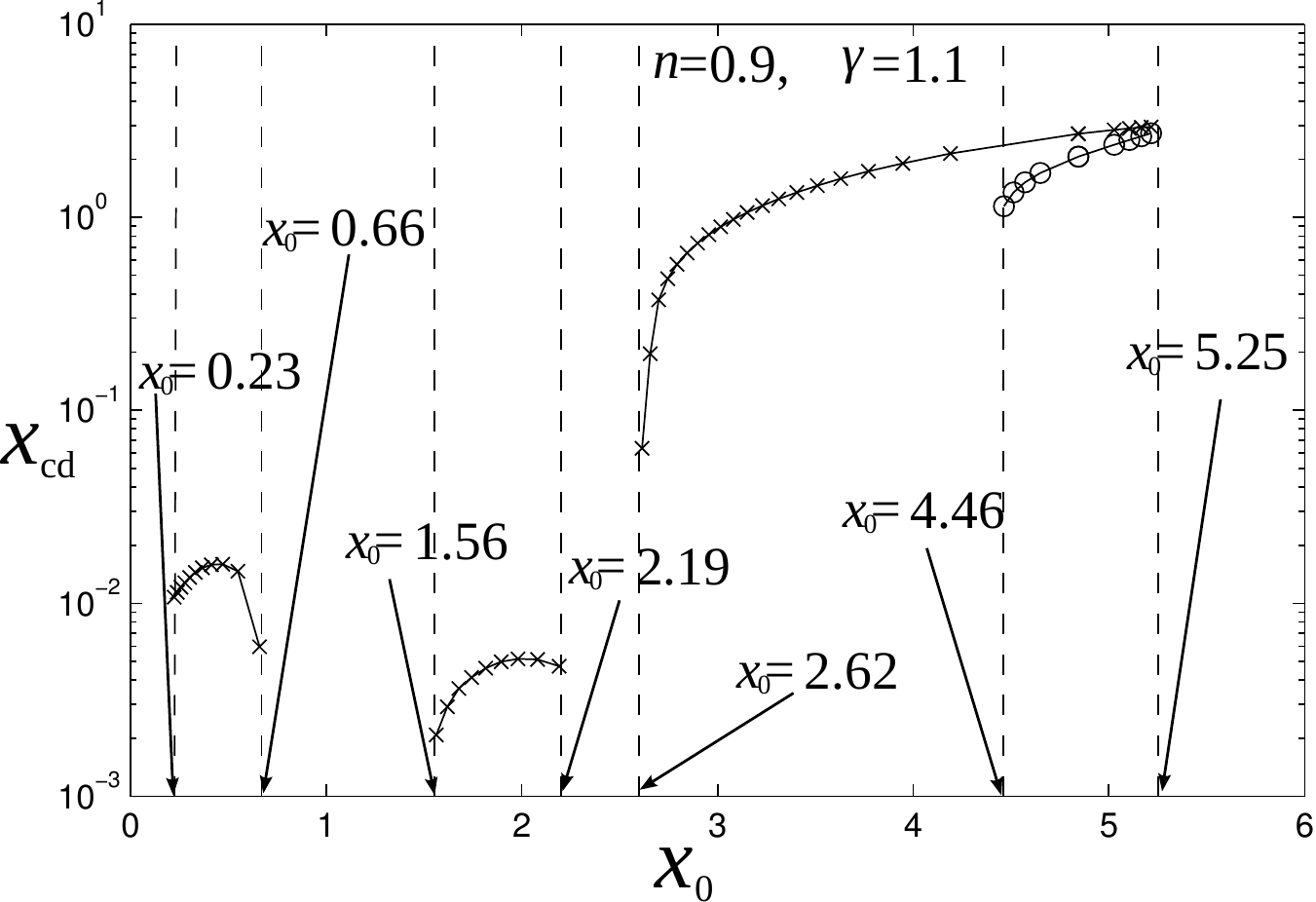}
  \caption{
The relation of $x_\text{cd}$ (where $\alpha_\text{cd}$
    and $v_\text{cd}$ can be determined) versus $x_0$
    with $n=0.9$ ($\gamma=1.1$).
The light solid curves marked along with crosses (``$\times$'')
    represent three branches of type-1 solutions, while those marked along
    with circles (``$\circ$'') represent one branch of type-2
    solutions.
Type-1 solutions are located in these intervals of $x_0$:
    $0.23<x_0<0.66$, $1.56<x_0<2.19$, $2.62<x_0<5.25$.
Type-2 solutions, on the other hand, are located in
    one interval of $x_0$, i.e. $4.46<x_0<5.25$.
The two rightmost branches of solutions
    in different type actually converge at $x_0\simeq 5.25$.
In fact, it is the very point where the square root in
    eq. \eqref{eq:V_PRIME} vanishes and thus the two
    types of solutions are identical.
}
    \label{fig:SCC_PD_0_9}
\end{figure}

Conventional polytropic void solutions can
 cross the SCC smoothly without shocks.
As shown in subsection \ref{sec:SCC}, we are able to
 construct this kind of eigensolutions with central
 voids in self-similar evolution of dynamic expansion.

While usual schemes always involve plotting a
 $\alpha-v$ phase diagram by integrate towards a
 chosen meeting point, this series of void solutions
 can also be constructed as follows without resorting
 to the $\alpha-v$ phase diagram.
First we choose a point $(x_0,\ v_0,\ \alpha_0)$ as the
 sonic critical point on the SCC which satisfies ODEs
 \eqref{eq:SCC_DENOMINATOR} and \eqref{eq:SCC_NUMERATOR}.
In fact, this represents a DOF. There,
 specific types (either type-1 or type-2, depending on which
 of the two $v'$ we take) of $v'$ and $\alpha'$ are calculated
 consequently by solving eq. \eqref{eq:SCC_DERIVATIVE}.
After choosing a proper $\delta<0$, we calculate $v$ and $\alpha$
 at $x=x_0+\delta<x_0$ and integrate inwards to see if it could
 reach the ZML; when this is fulfilled, let $\delta>0$ to obtain
 the reduced variables $v$ and $\alpha$ at $x=x_0+\delta>x_0$,
 we then integrate outwards to obtain a global solution for the
 self-similar dynamic evolution of an expanding central void.

In general, the SCC and ZML together enclose an area in
 the figure for presenting $v$ versus $x$ profiles.
Obviously, it is necessary for a void solution to have part
 of itself inside this area since it should touch the ZML.
Therefore, there is a necessary condition for a solution
 crossing the SCC smoothly to reach the ZML as
%  {\bf for $x<x_0$, inequality below
%   should be the other way around?}
\begin{equation}
  v'<\dfrac{\mbox{d}v_0}{\mbox{d}x_0}\ ,
\end{equation}
  where $v'$ carries the same meaning as in
  eq. \eqref{eq:SCC_DERIVATIVE}, of a specific
  type (either type-1 or type-2).
If this inequality is not satisfied along the whole SCC, we would
  expect no void solution that crosses the SCC smoothly of this type.
It is straightforward to derive from eqs. \eqref{eq:SCC_SOLUTIONS}
  and \eqref{eq:SCC_DERIVATIVE} that the necessary condition for the
  existence of type-2 void solution (with smooth behaviour across
  the SCC) is $n>0.840$ at segment 1 of SCC (for the definition
  of ``segment 1", see eq. \eqref{eq:SCC_SOLUTIONS} as well as
  related discussions and definitions).
%  {\bf What is this?}
By extensive numerical investigations, we find no type-2
  solutions with $x_0$ less than $50$ for $n=0.67$,
%  {\bf Have you tried other values of $n$?}
%  {\it In the mean time,
whereas for $n=0.9$, both types of void solutions exist.
%  }

As examples, we present several solutions ($n=0.67$ and $n=0.9$)
  crossing the SCC smoothly in Figs. \ref{fig:SCC_T1_0_67} and
  \ref{fig:SCC_0_9}, with the most important relevant parameter
  $x_0$ being chosen as $4.277$, $5.010$ and $6.084$ for Model
  S1, S2 and S3, respectively.
All those with $n=0.67$ are type-1 void solutions, while
  there are both types of void solutions for $n=0.9$.
Corresponding diagrams for the relation of $x_\text{cd}$
  versus $x_0$ for $n=0.67$ and $n=0.9$ are displayed in Figs.
  \ref{fig:SCC_PD_0_67} and \ref{fig:SCC_PD_0_9}, respectively.

\subsection{
 %Conventional polytropic
   Expanding void solutions with shocks}

\begin{figure}
  \includegraphics[bb=0 0 330 480, width=70mm]
  {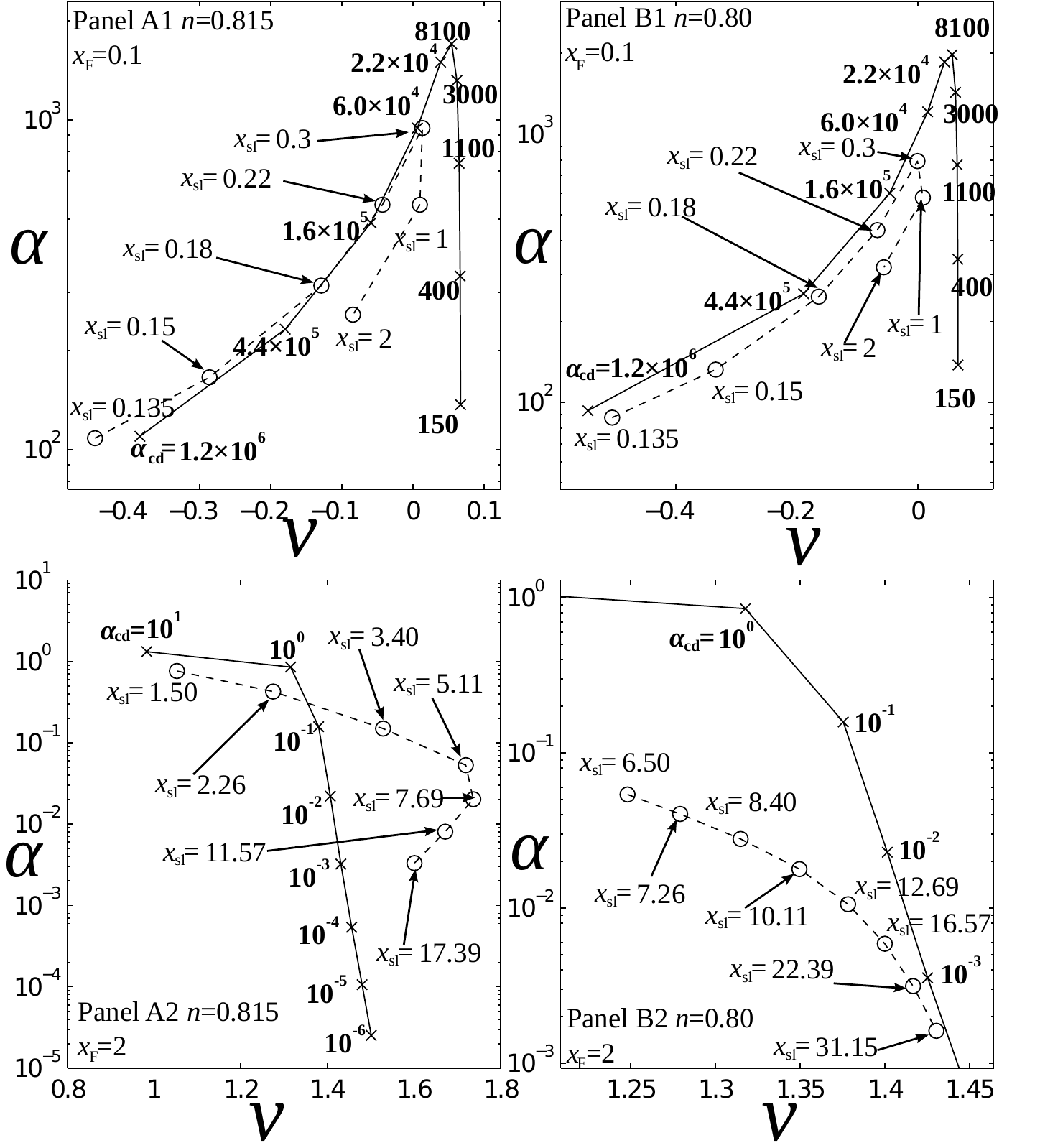}
  \caption{Presentation for the detachment (i.e. no cross-over)
    of the two phase curves
    in the phase diagram when $n$ decreases to $0.80$ (for the
    possible presence of shock solutions with EWCS envelopes).
This detachment does not occur even when $n$ is only slightly
    greater than $0.80$ (say, $0.803$).
To be specific, we compare $n=0.80$ cases
    with $n=0.815$ cases in this presentation.
In all four panels, dashed curves marked with
    circles (``$\circ$") are the phase curves
    representing different $x_\text{s1}$.
The solid curve marked with crosses (``$\times$")
    are the phase curves representing different $\alpha_\text{cd}$
    and $x_\text{cd}\rightarrow 0^+$: these curves are the inner
    envelope of the ``phase nets" with various combinations of
    $\alpha_\text{cd}$ and $x_\text{cd}$.
Panels A1 and B1 illustrate the change of the phase
    curves in the ``upper left region" (mainly for
    $x_\text{s1}<x_\text{e}$) while $n$ decreases
    from $0.815$ (Panel A1) to $0.80$ (Panel B1).
Panels A2 and B2, meanwhile, present this change in the ``lower
    right region" (for $x_\text{s1}>x_\text{e}$) in the phase diagram.
The detachment of the two phase curves (hence the detachment
    of the circled phase curve from the inner envelope of
    the ``phase net") is apparent from this presentation.
%    {\bf Need an overall clarification.}
}
    \label{fig:EWCS_PD_COMPARE}
\end{figure}

% Two procedures -- integrating inwards from a relatively large $x$
% with asymptotic condition given by
% eqs.\eqref{eq:ASYMPTOTIC_SOLUTIONS} by specifying relevant
% parameters and the non-dimensional location of shock ($x_\text{s}$)
% on the upstream side, or integrating outwards by giving $x$ (hence
% also $v$) and $\alpha$ at the contact discontinuity and $x_\text{s}$
% on the downstream side -- are both acceptable.

Shocks are another way by which solutions
 can go across the sonic critical surface.
There is an extra DOF for void solutions with shocks:
 the shock location in the self-similar expansion.
We have applied two types of numerical schemes
 to construct void solutions with shocks.
The following are the general outlines of these
 two procedures, perhaps with minor modifications
 in dealing with specific integrations.

One computation procedure is to integrate inwards. For solutions
  with vanishing $|v|$ and $\alpha$ at $x\rightarrow\infty$
  under consideration, we can use asymptotic solution
  \eqref{eq:ASYMPTOTIC_SOLUTIONS}
%  {\bf Extra $B$ term}
  to determine the initial value of integration
  at an $x$ that is large enough (say, $x=30$).
We also choose a self-similar shock location
  in the upstream flow at $x_{s1}$.
Then we integrate inwards numerically using the standard
  fourth-order Runge-Kutta scheme, apply shock jump
  conditions \eqref{eq:SHOCK_SOLUTIONS} in self-similar
  form at $x=x_\text{s1}$, and continue to integrate
  inwards from $x_{s2}$ and see if the solution would
  gradually approach and eventually touch the ZML.
When such a solution touches the ZML, a global void
  solution with shock can then be readily constructed.
Meanwhile, values of dimensionless variables at
  the contact discontinuity, $x_\text{cd}$
  ($v_\text{cd}=nx_\text{cd}$) and $\alpha_\text{cd}$,
  are determined in a consistent manner.

The other computation procedure is to integrate outwards. Choose
  the self-similar location of void boundary $x_\text{cd}$ and
  reduced mass density $\alpha_\text{cd}$ there, and decide
  the location $x_\text{s2}$ of shock on the downstream side.
Then integration is taken outwards to a relatively large value of
  $x$ with respect of the jump at the shock by applying shock
  conditions \eqref{eq:SHOCK_SOLUTIONS} (note the
  commutation symmetry with regard to subscripts 1 and 2
  discussed in subsection \ref{sec:SHOCK}).
The mass and speed parameters $A$ and $B$ to characterize the
  asymptotic solution behaviours at large $x$ can also be
  evaluated with the value of $v(x)$ and $\alpha(x)$ at a very
  large $x$ by using eq. \eqref{eq:ASYMPTOTIC_SOLUTIONS}.

Practically, there is not much difference between the two
  kinds of void solution construction procedures except
  when parameters chosen in highly ``sensitive" regimes.
We have applied both numerical integration procedures as
  cross-checks of each other for the reliability of our
  void solutions.

\subsubsection{Expanding void solutions with EWCS envelope}

\begin{figure}
  \includegraphics[bb=0 0 330 505, width=70mm]
  {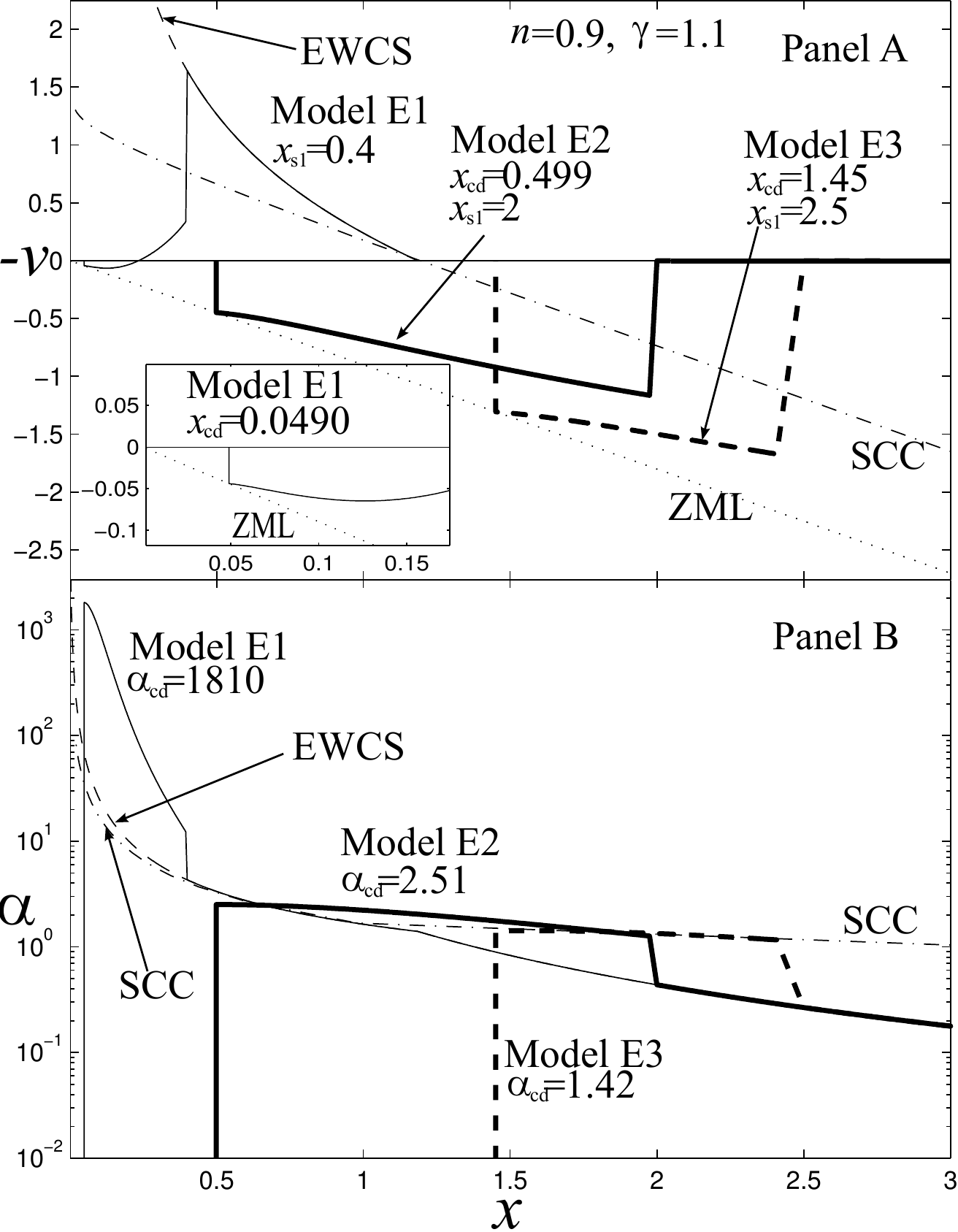}
  \caption{Void solutions with shocks and EWCS envelopes;
Panel A for the negative reduced radial velocity $-v(x)$
  versus $x$ profiles and Panel B for the reduced mass
  density $\alpha(x)$ versus $x$ profiles.
These conventional polytropic solutions have $n=0.9$
  ($\gamma=1.1$).
The SCCs in both panels are presented by dash-dotted curves,
   while the ZML in Panel A is shown by dotted line.
Models E1 (``E'' for ``EWCS''), E2 and E3 are shown
   by light solid curve, heavy solid curve and heavy
   dashed curve, respectively.
The envelopes beyond the shock of Models E2 and E3
   are the same: SPS (thus $x_\text{s1}>x_\text{e}$)
   solution, but the locations of the shocks are different.
Model E1, on the other hand, has $x_\text{s1}<x_\text{e}$. }
    \label{fig:EWCS_PROFILES}
\end{figure}

The outer part of EWCS is identical to the static
% singular polytropic sphere
 SPS solution, namely \citep[e.g.][with
 $n+\gamma=2$]{1988ApJ...326..527S}
\begin{equation}
  v = 0\ ,\qquad\qquad
  \alpha =\left[\dfrac{2(2-n)(3n-2)}{n^2}\right]^{1/n}x^{-2/n}\ ,
\end{equation}
beyond the expansion wave radius ($x_\text{e}$ for the similarity
 location of this radius), i.e. $x>x_\text{e}$.
For $n=1$, this type of polytropic solution becomes
 the outer part of a singular isothermal sphere
 \cite[SIS; e.g.][2010]{springerlink:10.1007/s10509-009-0044-4}.
The asymptotic behaviour of this kind of
 EWCS at small $x$ is given by
 \cite[e.g.][]{1988ApJ...326..527S,2006MNRAS.372..885L}
\begin{equation}
  v=-\left[\dfrac{2m(0)}{(3n-2)x}\right]^{1/2}\ ,\qquad
  \alpha=\left[\dfrac{(3n-2)m(0)}{2x^3}\right]^{1/2}\ ,
\end{equation}
where constant $m(0)$ represents the central mass point.
%For $n=1$, this type
% of polytropic solution becomes the singular isothermal sphere
% \cite[SIS; e.g.][]{springerlink:10.1007/s10509-009-0044-4}.
% {\bf Please clarify and be precise. }

We have obtained shock solutions with
 central void and EWCS envelope.
When a specific envelope is prescribed, there
 is only one DOF, viz., the shock location.
The technique of $v$ versus $\alpha$ ``phase net'', a
 variation of $v$ versus $\alpha$ phase diagram scheme
%(the phase diagram scheme was introduced by
 [\citet{1977ApJ...218..834H} and the ``phase net" was
 developed in \citet[][2010]{springerlink:10.1007/s10509-009-0044-4}],
 is applicable because of the similar situation.
Steps of the scheme are summarized below.
 We first obtain the EWCS envelope
 by numerical integration. Choose
 a meeting point $x_\text{F}$ and the similarity upstream
 location of the shock $x_\text{s1}$. With
 shock jump conditions applied for connecting $x_\text{s1}$
 and $x_\text{s2}$, we numerically integrate from $x_\text{s2}$
 to $x_\text{F}$ and record the value of $v$ and $\alpha$ there.
Select a series of $x_\text{s1}$ values, get a series of
 $(v,\ \alpha)$ pairs at $x_\text{F}$ and use these $(v,\ \alpha)$
 pairs to plot a curve, on which different points correspond to
 different values of $x_\text{s1}$, in the phase diagram.
Then we select a series of $x_\text{cd}$ and $\alpha_\text{cd}$
 in pairs and do similar things: integrate towards $x_\text{F}$
 and get a sequence of curves or a ``net" [i.e., the so-called
 ``phase net" in \citet{springerlink:10.1007/s10509-009-0044-4}]
 in the phase diagram corresponding to different
 $x_\text{cd}$ and $\alpha_\text{cd}$ in pairs.

Applying this scheme, we have explored various void
 solutions with static polytropic envelopes.
We find numerically that when the scaling
 index $n$ is less than $0.80$, there appears to be no
 such type of solutions with one shock.
We realize for any specific $n$ that the curve
 representing $x_\text{s1}$ is actually bounded in
 the phase diagram and ``shrinks" as $n$ increases.
Meanwhile, the ``net" with different $x_\text{cd}$ and
 $\alpha_\text{cd}$ pairs has an inner envelope (actually
 this envelope is outlined by the curve with different
 $\alpha_\text{cd}$ and $x_\text{cd}\rightarrow 0^+$).
When $n$ becomes greater than $0.80$, there are two regions in the
 diagram where the $x_\text{s1}$-curve intersects the ``net": one
 corresponds to the solutions with $x_\text{s1}<x_\text{e}$, and
 the other, $x_\text{s1}>x_\text{e}$ (in the phase diagram, these
 two regions are clearly separated).
 %{\bf What?}
However, when $n$ becomes less than $0.80$, the
 $x_\text{s1}$-curve detaches from the net and neither
 of the two intersecting areas can still exist.
We have tried various values of $x_\text{F}$ from $0.01$
 to $20$ and only find the same threshold of $n=0.80$.
%{\bf Can you somehow demonstrate this
%critical $n$ value analytically?}

We shall refer to the region with $x_\text{s1}<x_\text{e}$ as the
  ``upper region" and that with $x_\text{s1}>x_\text{e}$ as the
  ``lower region" in the caption of Fig.
  \ref{fig:EWCS_PD_COMPARE} because in a more complete
  phase diagram, the ``upper region" always appears on
  the upper-left corner while the ``lower region"  is always on
  the lower-right corner.
%If the reader find those statements confusing, a reference to
These terminologies can be readily clarified by referring to
  figure 7 of \citet{springerlink:10.1007/s10509-009-0044-4}
  for an isothermal case.
%) will illustrate the notation of  ``upper" and ``lower" clearly.
%
In addition, if there is no EWCS-envelope void solution (with shock),
 there will be no breeze solution with a void and a shock either.
We have found that if $0<A<A_\text{e}$ and $B=0$ (hence $v>0$
 at large $x$), the curve for $x_\text{s1}$ will ``shrink"
 even more, further preventing $x_\text{s1}$ curve from
 intersecting the ``net".
Furthermore, the decrease of $B$ parameter also leads to a
 ``shrink'' of $x_\text{s1}$ curve, therefore contraction
 and inflow solutions become much more difficult to exist.

This detachment in the phase diagram related to the
 scaling index $n=0.8$ is illustrated by
 examples in Fig. \ref{fig:EWCS_PD_COMPARE}.
Note that this separation by $n\lsim 0.8$ is valid
 for conventional polytropic cases only; there is
 no such separation for general polytropic cases
 as shown in \citet {2008MNRAS.390.1619H}.
% {\bf Please clarify more specifically!}
Note also that void solutions with EWCS envelope and shock may
 still exist if the solutions cross the sonic critical surface
 for more than once (crossing smoothly or by shock discontinuity).

Several void solution examples of this type are presented in
 Fig. \ref {fig:EWCS_PROFILES} with $n=0.9$ (thus $\gamma=1.1$)
 and $x_\text{s1}$ being $0.4$, $2$, and $2.5$, respectively.

\subsubsection{Void solutions with various dynamic
  envelopes: breeze, contraction, outflow and inflow}

% {\bf What is this?}
With relevant parameters summarized in Table
 \ref{table:PARAMETER_WITH_SHOCK}, a series
 of self-similar void solutions are obtained\footnote{The
 results in Table 1 are obtained with the inclusion of
 $-(n-1)B^2x^{1-2/n}/n$ term (Lou \& Shi 2011 in preparation).
The values of mass and velocity parameters $A$ and $B$,
 especially those with $n=0.67$, have already been modified as
 compared with those without including this term.
In contrast, $n=0.9$ cases are almost not influenced
 by the inclusion of this $B^2$ term.}
 with shocks and $n=0.67$ (values of $n$ are
 to be discussed in subsection \ref{sec:EOS}).
% by integrating inwards with the standard fourth-order Runge-Kutta
% scheme, with a step length of $10^{-5}$ to minimize overall errors
% regarding the accuracy of variables -- 16 effective digits -- stored
% in the computer. Results have already been also checked by
% integrating outwards.

\begin{figure}
  \includegraphics[bb=0 0 330 505, width=70mm]
  {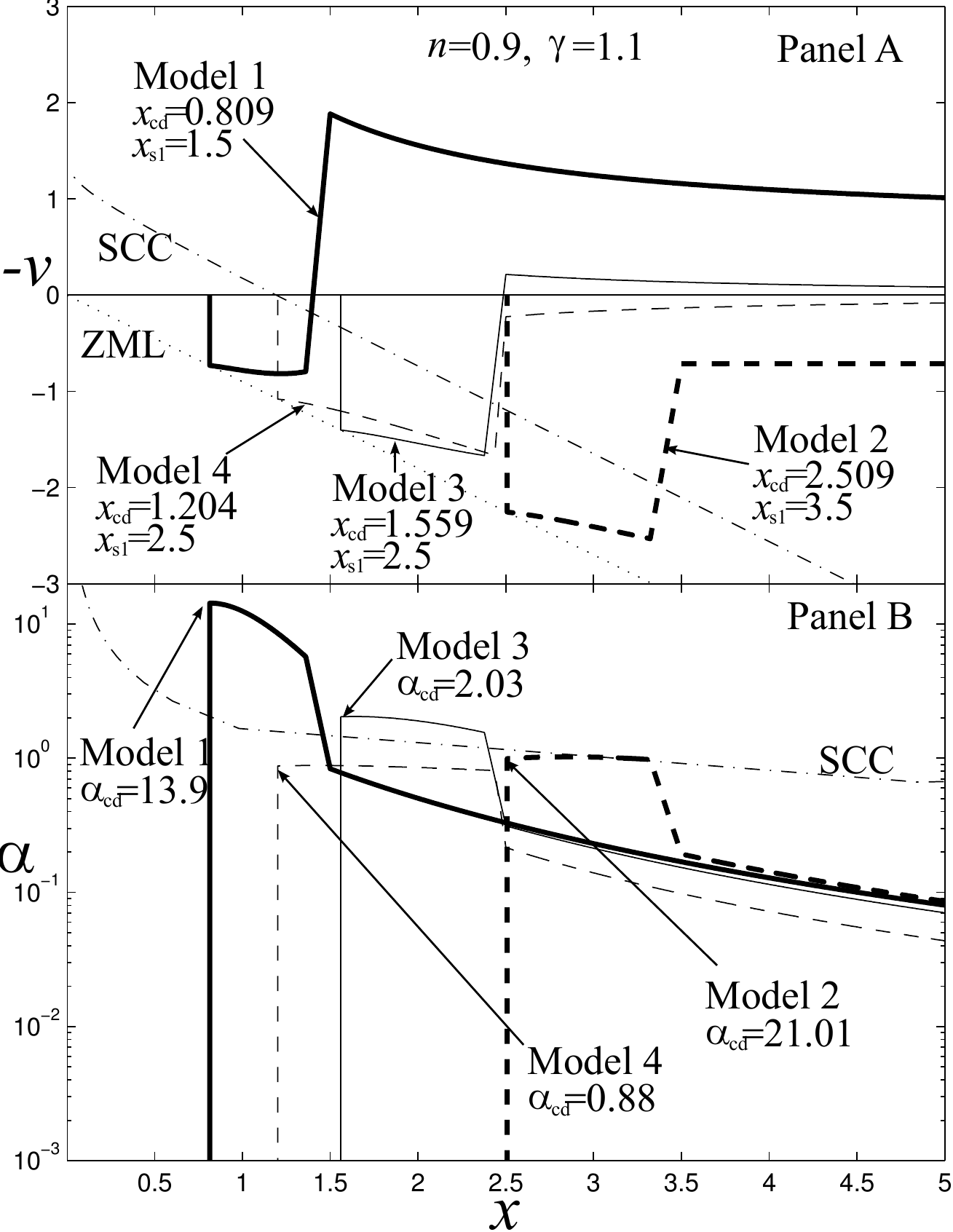}
  \caption{Presentation of negative reduced radial flow velocity
$-v(x)$ versus $x$ (Panel A) and reduced mass density $\alpha(x)$
  versus $x$ (Panel B) for shock void solutions with inflow
  (heavy solid curve), outflow (heavy dashed curve), contraction
  (light dashed curve) and breeze (light solid curve) envelopes
  specified by Models 1 through 4 in Table
  \ref{table:PARAMETER_WITH_SHOCK}, respectively.
The SCCs in both Panels A and B are shown by dash-dotted curves;
  the ZML is represented by the light dotted curve. }
  \label{fig:B_C_I_O_0_90}
\end{figure}

\begin{figure}
  \includegraphics[bb=0 0 330 498, width=70mm]{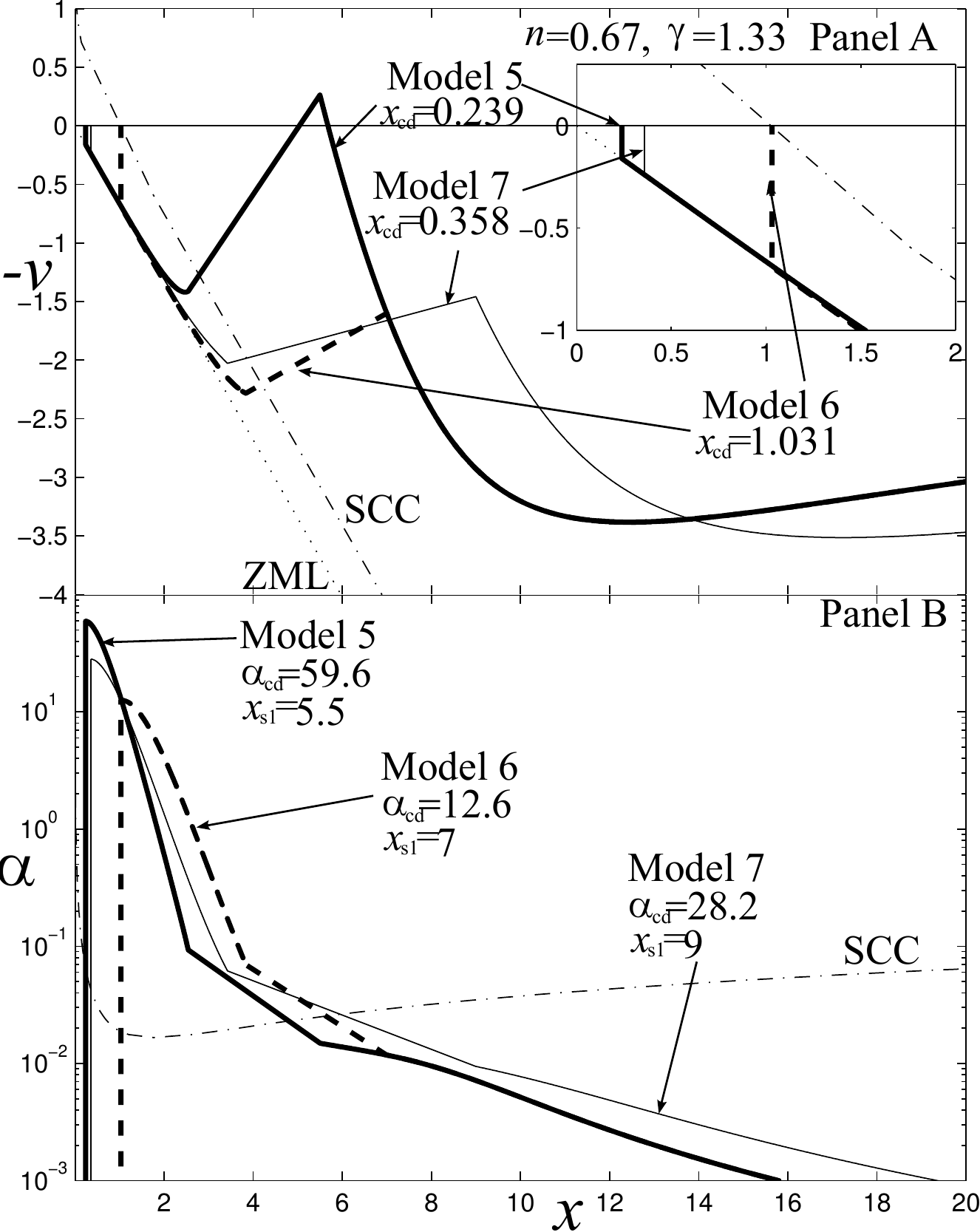}
  \caption{
  %{\bf Please use larger fonts for labels!!}
Presentation of negative reduced radial flow velocity $-v(x)$
   versus $x$ (Panel A) and reduced mass density $\alpha(x)$
   versus $x$ (Panel B) for outflow-envelope solutions specified
   by Models $5-7$ in Table \ref{table:PARAMETER_WITH_SHOCK}.
The dash-dotted curves in both panels are the SCCs. The light
   dotted line in Panel A is the ZML. Among the three solutions,
Models 5, 6 and 7 are illustrated with heavy solid curve, heavy
   dashed curve and light solid curve, respectively.
The small inset at the upper-right corner of Panel A is a zoom
   of solution curves near the origin $x=0$ and $-v=0$. }
    \label{fig:O_0_67}
\end{figure}

Models 1 through 4 are four kinds of void solutions: those
 that have breeze, contraction, outflow and inflow envelopes,
 respectively, all with $n=0.90>0.80$.
Their solutions are illustrated in
%a different plotting
 Fig. \ref{fig:B_C_I_O_0_90} to show in non-dimensional form
 the profiles of reduced radial velocity $v(x)$ and reduced mass
 density $\alpha(x)$ (definitions of these four kinds of envelopes
 are given in subsection \ref{sec:DYNAMIC_EQS_SIMILARITY_TRANS}).
Models 5 through 7 are void solutions with outflow envelope
 beyond the shock with $n=0.67<0.80$, which we shall
 elaborate for further applications.
These solutions are displayed in Fig. \ref{fig:O_0_67}.

In numerical explorations for cases of $n=0.67$, we
 have found that even for strong outflows, there are
 impressive trends to fall inwards at some intervals
 of $x$: this is clearly the results of self-gravity.
This point is unambiguously illustrated by Models 5 through 7, and
 other models not shown here also appear similarly in this aspect.
We shall further discuss this feature in subsection
 \ref{sec:DIMENSIONAL_MODEL}.

\begin{table}
\centering \caption{
Parameters specifying conventional polytropic
  void solutions with shocks.
The first four models all have $n=0.9$ while
  the last three all have $n=0.67$.
These solutions are constructed by
  integrating inwards and we confirmed
  their correctness by integrating
  outwards within the tolerance
  of numerical errors. }
\begin{tabular}{cccccccc}
  \\
  \hline
  Model&$n$&$A$&$B$&$x_\text{s1}$&$\eta_k$&$x_\text{cd}$&$\alpha_\text{cd}$\\
  \hline
  1 & 0.9  & 3     &  $-1$  &1.5 &  0.826& 0.809 &13.9 \\
  2 & 0.9  & 3     &    1   &3.5 &  0.902& 2.509 &1.01 \\
  3 & 0.9  & 2.542 &    0   &2.5 &  0.905& 1.559 &2.03 \\
  4 & 0.9  & 1.542 &    0   &2.5 &  0.950& 1.204 &0.88 \\
  5 & 0.67 & 2.43  &   13.6 &5.5 &  0.214& 0.239 &59.6 \\
  6 & 0.67 & 2.25  &   13.2 &7   &  0.300& 1.031 &12.6 \\
  7 & 0.67 & 4.50  &   16.7 &9   &  0.145& 0.358 &28.2 \\
  \hline
\end{tabular}
\label{table:PARAMETER_WITH_SHOCK}
%The results in this table are produced with
%the extra $-(n-1)B^2x^{1-2/n}/n$ term.
%The values of $A$ and $B$, especially those with
%  $n=0.67$, have already been modified.
%However, $n=0.9$ cases are almost not
%  influenced by the extra $B^2$ term.
\end{table}

\section{Astrophysical Applications}
\label{sec:APPLICATIONS}

There are several astrophysical situations where hot tenuous
 bubbles exist, shaped up by pertinent physical mechanisms.
Here, we discuss cases that might be capable of describing the
 evolution of supernova at an early stage (the so-called
 ``optically thick" stage), during which the predominant
 driving power inside the stellar envelope could possibly be
 the pressure of some extremely relativistic particles such
 as neutrinos and photons trapped inside a central cavity.
Some necessary aspects need to be addressed
 before outlining our model scenario.

\subsection{Formation of a central bubble or cavity}

Research works have hinted at a scenario that a cavity
 can be formed surrounding the centre of a supernova
 during the initial phase.
As concluded by \citet{RevModPhys.62.801}, accelerating infall
 of substances is inhibited by powerful neutrino pressure
 from the neutrinosphere before neutrinos decouple from the gas
 and escape; moreover, a rebound shock may be revived when
 the intense neutrino flux is absorbed by surrounding nuclear
 matters at a radius $r\sim 100\mbox{ km}$ or more.
Such a mechanism can drive materials around the core outwards
 and shape up a rarified bubble or cavity, which is clearly
 illustrated in figure 2 of \citet{1985ApJ...295...14B}.
This perspective was further strengthened by simulations
 of \citet[][]{1989A&AS...78..375J,1989A&A...224...49J}
 and Janka \& M\"uller (1996),
 %1996A&A...306..167J},
 revealing the formation of a bubble around the centre of a
 supernova, filled with intense electromagnetic radiation
 field and other relativistic materials.

\subsection{Pressure balance, EOS, and the central power}
\label{sec:EOS}

While it is allowed to be discontinuous in mass densities
 and temperatures across the contact discontinuity surface,
 a pressure balance across this contact surface should be
 maintained to fulfill the necessary mechanical requirement.
Here, we discuss consequences for the conventional polytropic
 EoS and some relevant aspects of the required pressure balance.

For an extremely relativistic degenerate or hot gas,
 statistical mechanics gives an adiabatic EoS
 \citep[e.g.][]{CALLEN_THERMODYNAMICS},
\begin{equation}
\label{eq:ADIABATIC_EQ_NEUTRINO}
  p \rho^{-4/3} = \mbox{constant}\ .
\end{equation}
%{\bf Define notations. }
With a spherical volume $V=4\pi r^3/3$, $\rho$ is the
 mass density, $p$ is the pressure, and the energy density
 is simply gotten by
 a multiplication of $c^2$ with $c$ being the speed of light.
%This EoS is especially true for the degenerate relativistic
% materials, whose thermal pressure depends almost solely on
% the mass density. {\bf Please clarify more specifically.}

Consider matters inside a highly rarified central cavity. For a
 temperature as high as $\sim 1-10\mbox{ MeV}/k_\text{B}$ or more,
 it is natural to expect substances (i.e. radiation field and
 products of electron-positron pair production processes)
 inside the cavity to be relativistic.
During the self-similar dynamic expansion of a central void,
 transformation \eqref{eq:SELF_SIMILAR_TRANSFORMATION}
 gives $r_\text{cd}\propto t^n$.
Assuming that the mass (hence energy) inside the void is to be
 homogeneous and conserved (adiabatic) during the self-similar
 expansion, thermal pressure $p$ is thus proportional to
% {\bf You mean $V^{-4/3}\propto r_\text{cd}^{-4}$?}
 $V^{-4/3}\propto r_\text{cd}^{-4}$ (for relativistic matters,
 the speed is almost the speed of light $c$; hence homogeneity
 would be a sensible approximation for a not very large scale,
 say $\lsim c\times 1\mbox{ s}$).
From those, we conclude that $p\propto t^{-4n}$ for
 relativistic matters inside a void at the boundary
 (i.e., the surface of contact discontinuity).

On the other hand in terms of the surrounding gas envelope in
 self-similar evolution, the thermal pressure just outside
 the contact surface is proportional to $t^{2n-4}$ by
 transformation \eqref{eq:SELF_SIMILAR_TRANSFORMATION};
a contact surface has constant $\alpha_\text{cd}$
 and $x_\text{cd}$ in a self-similar dynamic evolution.
There should be a sustained pressure balance across the contact
 surface, requiring at least the same time-dependence of
 pressures on both sides of a contact discontinuity.
With this consideration alone, we would require $n=2/3$
 from $2n-4=4n$ (i.e. $\gamma=4/3$ from $n+\gamma=2$).

%In spite that adiabatic (or mass conserving) condition for
%  substances inside the void would be generally preferred,
We shall not consider the exact case of $n=2/3$ for a
 homologous flow as done by \citet{1980ApJ...238..991G},
 \citet{1983ApJ...265.1047Y}, and \citet{2008MNRAS.384..611L}.
%{\bf Need to modify this statement.}
Instead, we invoke a physically more plausible
  condition $n=2/3+\epsilon$ with
  $\epsilon\gsim 0$ being a real number slightly greater
  than zero (e.g., $n=0.67$) to describe a small deviation from
  an adiabatic process for extremely relativistic matters inside
  a cavity, with a tacit assumption that the central compact
  remnant (e.g. a nascent neutron star or a nascent stellar
  mass black hole) continues to input energy into the ``void"
  with a time-dependent rate, usually a decreasing one.
%      instillation {\bf Meaning here?}

Simple thermodynamic and mechanical considerations yield that an
 $\epsilon$ added to $n$ reflects some physical mechanisms adding
 more energy into a void during its dynamic expansion, viz.
\begin{equation}
  \mbox{d}Q\simeq 3\epsilon p\mbox{d}V\ ,
\end{equation}
where $Q$ is the amount of heat input into a
  void by some mechanisms as shown in Appendix B.

By transformation \eqref{eq:SELF_SIMILAR_TRANSFORMATION},
 we readily derive
\begin{equation}
  \label{eq:ENERGY_INPUT_RATE}
  \dfrac{\mbox{d}Q}{\mbox{d}t}=3\epsilon\left(\dfrac{2}{3}+
    \epsilon\right)\alpha_\text{cd}^{4/3-\epsilon}x_\text{cd}^3
    \dfrac{k^{5/2}t^{5\epsilon-5/3}}{G}\ ,
\end{equation}
where $G$ is the gravitational constant and $k$
  is the same sound parameter in transformation
  \eqref{eq:SELF_SIMILAR_TRANSFORMATION}, and the subscript
  ``cd" denotes quantities related to the contact discontinuity.
The sound parameter $k$ should be evaluated with in terms
  of dimensional values such as $\rho_\text{cd}$, which
  depends on specific applications of our void model.

We do not yet know the specific form(s) of energy release
 from the central compact object into the surrounding cavity.
We consider simple cases as examples of model consideration.

%***************************************
There is a sensible physical possibility that the energy
  input comes mainly from the thermal radiation from the
  collapsing central compact object.
As is conventional in study of neutron star cooling process,
  an ``effective" temperature $T_\text{e}$ is often introduced
  to indicate the ability of photon radiation
  \citep[e.g.][]{2004ARA&A..42..169Y,2006NuPhA.777..497P}.
%  which we use $T_\text{e}$ to denote.
Note that even though neutrino cooling is much more important
  than photon cooling during this epoch, photon radiation is
  much more important for a sustained expansion of central
  void (see subsection \ref{sec:MOMENTUM_TRANSFER_ESTIMATION}
  for details).
With this $T_\text{e}$ specified, the radiation
  energy flux input into the bubble is given by
\begin{equation}
  \label{eq:ENERGY_INPUT_THERMAL}
  \dfrac{\mbox{d}Q}{\mbox{d}t}=4\pi r_\text{n}^2\sigma
  T_\text{e}^4\ ,
\end{equation}
  where $r_\text{n}$ is the radius of the proto-neutron
  star and $\sigma=5.6705\times 10^{-5}\mbox{ erg cm}^{-2}
  \mbox{ s}^{-1}\mbox{ K}^{-4}$ is the Stefan-Boltzmann constant.
Combining eqs. \eqref{eq:ENERGY_INPUT_RATE} and
  \eqref{eq:ENERGY_INPUT_THERMAL} with $n=2/3+\epsilon$,
  we obtain a scaling of $T_\text{e}$ as a function of
  $t$ in the form of
\begin{equation}
  T_\text{e}\propto t^{5(n-1)/4}\ .
\end{equation}
If simplified calculations of \citet{2006NuPhA.777..497P}
  hold even in the early evolution phase of a proto-neutron
  star, we obtain approximately for a ``slow" (``standard")
  cooling process, $T_\text{e}\propto t^{-1/12}$ and thus
  $n=14/15\simeq 0.933$, while for a ``fast" cooling process,
  $T_\text{e}\propto t^{-1/8}$ and thus $n=0.9$.

As an example, let us approximately calculate the amount of
  energy input rate with respect to the model applied in our
  scenario for SN1993J in subsection \ref{sec:SN1993J}.
The dimensionless model applied there has $x_\text{cd}=2$ and
  $\alpha_\text{cd}=10^{-3}$ with $n=0.933$ (in addition, the
  relevant $x_\text{s2}$ is $3.3$).
The dimensional counterpart is taken as what is specified
  in Fig. \ref {fig:SN1993J_INITIAL} at $t=11\mbox{ s}$.
We then have the sound parameter
  $k=1.584\times 10^{18}\mbox{ c.g.s. unit }$
  and thus from eq. \eqref{eq:ENERGY_INPUT_RATE}
\begin{equation}
  \dfrac{\mbox{d}Q}{\mbox{d}t}\simeq
  8\times 10^{49}\mbox{ erg s}^{-1}\ .
\end{equation}
A thermal luminosity as high as this would be possible in
  a very early stage in terms of pertinent calculations
  of \citet{2006NuPhA.777..497P} in their figure 14.
For $r_\text{n}=10\mbox{ km}$, we obtain from
  eq. \eqref{eq:ENERGY_INPUT_THERMAL} an effective temperature
  $T_\text{e}\simeq 1.8\times 10^{10}\mbox{ K}$ at that time,
  corresponding to a very early epoch of a proto-neutron star
  (see again figure 14 in Page et al. 2006 for a high surface
  temperature of a strange star).
This temperature $T_\text{e}$ is much higher than the radiation
  temperature in the bubble (see Fig. \ref{fig:SN1993J_INITIAL}).
Moreover, the temperature in the bubble drops at a rate
  proportional to $t^{n/2-1}=t^{-0.534}$, which is much
  faster than $T_\text{e}\propto t^{-1/12}$.
This can be seen from the pressure confining the central
  bubble $p=p_\text{cd}\propto t^{2n-4}$ [see also
  transformation \eqref {eq:SELF_SIMILAR_TRANSFORMATION}]
  and $p\propto T^4$ for radiation [see eq. (3.53) in
  \citet[][]{CALLEN_THERMODYNAMICS}].
%Therefore the energy that ``comes back" to the neutron star
%  surface from the radiation field inside the bubble can be
%  simply ignored.
%**********************************************

%There is a sensible physical possibility that the heat energy
% input comes mainly from radiations/emissions from the
% central compact object for a certain period of time.
%For example,
To continue the above consideration, we may presume
 approximately a black body emission spectrum
 from an isothermal proto-neutron star.
If the heat capacity at constant volume of
 a proto-neutron star depends on the temperature
 in a power-law form, viz. $C_V\propto T^\xi$
 with $\xi$ being an exponent, we have
\begin{equation}
  T_\text{n}^\xi\mbox{d}T_\text{n}\propto\mbox{d}U_\text{n}
  \propto -T_\text{n}^4\mbox{d}t\ ,
\end{equation}
where $U_\text{n}$ is the internal
  energy of the proto-neutron star.
  From that, we immediately derive the proportional
  relation $T_\text{n}\propto t^{1/(\xi-3)}$ or
\begin{equation}
  \dfrac{\mbox{d}Q}{\mbox{d}t}=
  -\dfrac{\mbox{d}U_\text{n}}{\mbox{d}t}
  \propto T_\text{n}^4\propto t^{4/(\xi-3)}\ .
\end{equation}
In reference to eq. \eqref{eq:ENERGY_INPUT_RATE}, we
  need to require $4/(\xi-3)=5\epsilon -5/3$, leading
  to the following relation between $n$ and $\xi$
\begin{equation}
  \label{eq:N_VS_XI}
  n=1+4/[5(\xi-3)]\ .
\end{equation}
Since for a degenerate Fermi gas, theoretical computations of
  heat capacity at constant volume $C_V$ gives the exponent
  $\xi$ range of $0<\xi<1$
  \citep[e.g.][from its figure 14.2]{GREINER_STAT}.
% \citep[e.g.][]{CALLEN_THERMODYNAMICS}.
It follows that scaling index $n$ would be
  restricted to the range $3/5<n<11/15$
  according to eq. \eqref{eq:N_VS_XI}.
Actually, we should require $n>2/3$ in our
  formulation for physical similarity solutions.
In other words, the mechanism attributing the energy
  input to the black body radiation from a central
  proto-neutron star to a surrounding hot bubble or
  cavity can only be responsible for the cases where
  inequality $2/3<n<11/15$ is satisfied.
%
%This consideration assumes that the entire proto-neutron
%  star is isothermal, while a more elaborate treatment
%  about heat transfer might make the radiation flux
%  lower than the isothermal case and thus permit
%  a larger value for $n$.
%
We emphasize that if for some reason, $\xi$ is allowed
 to be negative, then the value of $n$ may approach 1.
For other physical mechanisms of energy input into the
  central cavity from a proto-neutron star, inequality
  $n<11/15$ may not be necessary.
For example, numerical simulations of \citet{2001ApJ...562..887T}
  have offered a neutrino luminosity proportional to $t^{-0.9}$
  (see their figure 9) after a SN explosion and this would
  correspond to $n=0.82>11/15$.

\subsection{Momentum transfer by scattering processes}
\label{sec:MOMENTUM_TRANSFER_ESTIMATION}

The initial rapid detachment of an out-flowing envelope from
 a collapsing compact core is primarily accomplished by
 a powerful neutrino pressure \citep[e.g.][]{RevModPhys.62.801}.
After hundreds of milliseconds, energetic neutrinos decouple
 from surrounding gas materials and escape, an important
 source of driving power for the bubble expansion vanishes.
%But the envelope is unlikely to slowdown its expansion so
% quickly, which maintains a considerable amount of energy
% consumption.

Even though the radiation (mainly involving trapped photons
 and electron-positron pairs produced by pair production) is
 relatively weaker in luminosity than neutrinos, scattering
 process, by which the momentum and energy are transferred,
 of photons by matters in the gas shell is nevertheless much
 more effective.

Meanwhile, previous studies have implied that the amount of energy
 needed to blow a massive stellar envelope up is roughly at the same
 magnitude ($\sim 10^{51}\mbox{ erg}$) as that radiated by photons
 in SN explosion (see \citealt{1996A&A...306..167J} for details;
 this estimate could also be derived from an integration of
 the energy spectrum in \citealt{1974ApJ_188_501C}).
These also suggest that trapped photon radiation may be a major
 driving power of further detachment between a collapsing core
 and an expanding massive envelope after neutrinos have already
 shaped an initial central cavity and decoupled from surrounding
 gas materials.

Such a model consideration is outlined in the following
 subsection, where we compare the contributions of
 neutrino flux and of photon radiation field.
We will examine the possibility of photon radiation field
 of being an important driving power to the expansion of
 a shocked massive envelope and a sustained central void
 expansion.

\subsubsection{Scattering of neutrinos by heavy nuclei}

Energetic neutrinos are important in the
 initiation and evolution of SN explosions
% in prevailing astrophysical theories
 \citep[e.g.][]{RevModPhys.62.801}.
Scattering process of neutrinos with extremely high
  density matters is treated as coherent scattering,
  with a scattering cross section $\Sigma_\nu$
  given by \citep[e.g.][]{1975ApJ...201..467T}
% in its eq. (52)
\begin{equation}
  \label{eq:NEUTRINO_CROSS_SECTION}
  \Sigma_\nu\sim 10^{-45}\mbox{ cm}^2 A^2
  \left(\dfrac{E_\nu}{m_\text{e}c^2}\right)^2 ,
\end{equation}
  where
%  $\Sigma_\nu$ is the scattering cross section,
  $A$ is the number of nucleons in one nucleus,
  $E_\nu$ is the neutrino energy, $m_\text{e}$
  is the electron mass and $c$ is the speed of light.
The electron rest mass energy $m_\text{e}c^2$ comes
  basically from the process
  $\text{n}\rightarrow\text{p}+\text{e}^-+\bar{\nu}_\text{e}$
  in a nuclear reaction.
Equation \eqref{eq:NEUTRINO_CROSS_SECTION} does not include
  $\mu$ neutrinos $\nu_\mu$ ($\bar{\nu}_\mu$) and $\tau$
  neutrinos $\nu_\tau$ ($\bar{\nu}_\tau$), because the
  latter two are much less significant as compared with
  electron neutrinos $\nu_\text{e}$ ($\bar{\nu}_\text{e}$).

For a highly degenerate dense stellar core, the scattering of
  neutrinos are so effective that a central ``neutrino bubble"
  on a spatial scale of $\sim 100\mbox{ km}$ is shaped up by the
  extremely intense neutrino flux released from the collapsed core
  \citep[e.g.][]{RevModPhys.62.801,THEORETICAL_ASTROPHYSICS}.
In this case, the mean free path of electron
  neutrinos $\lambda_\nu$ is given by
  \citep[e.g.][]{THEORETICAL_ASTROPHYSICS}
\begin{equation}
  \lambda_\nu \simeq 3\times 10^4\mbox{ cm}
  \left(\dfrac{\bar{A}}{56}\right)^{-2}
  \left(\dfrac{\rho}{10^{12}\mbox{ g cm}^{-3}}\right)^{-5/3}\ ,
\end{equation}
where $\bar{A}$ is the average number of nucleons per nucleus.
% {\bf Define notations!}
On the other hand, the shaping of the neutrino bubble causes
  the matters surrounding the core to expand significantly,
  making them less dense and non-degenerate gradually (see
  Appendix \ref{sec:DIFFUSION}).
For a non-degenerate gas envelope, $\lambda_\nu$ is estimated by
\begin{equation}
  \lambda_\nu\simeq\dfrac{\bar{A}m_\text{p}}{\Sigma_\nu\rho}\sim
  0.5R_\odot\bar{A}^{-1}\left(\dfrac{E_\nu}{1\mbox{ MeV}}\right)^{-2}
  \left(\dfrac{\rho}{10^{10}\mbox{ g cm}^{-3}}\right)^{-1} .
\end{equation}
This $\lambda_\nu$ would be too long to accomplish an effective
  momentum transfer for a mass density less than that of nuclear matters.
Therefore we expect the material shell to be transparent for
  neutrinos shortly after a ``neutrino bubble" being shaped up.

\subsubsection{Scattering of photons by charged particles}
\label{sec:PHOTON_SCATTERING}

%{\it
Classical and semi-classical theories yield almost the same result for
  the scattering cross section $\Sigma_\text{ph}$ between a photon and
  a charged particle, usually referred to as the ``Thomson cross section''
  given by
%  which we use $\Sigma_\text{ph}$ to denote.
%This $\Sigma_\text{ph}$ is given by
  \citep[e.g.][]{JACKSON_CED}
%}
% eq.(14.126) {\bf Define notations!!}
\begin{equation}
  \label{eq:THOMSON_SECTION}
  \Sigma_{\text{ph}}=\dfrac{ 8 \pi }{ 3 }
  \left(\dfrac{e^2}{m c^2}
  \right)^2\ ,
\end{equation}
where $e=4.803\times 10^{-10}\mbox{ e.s.u.}$
  is the electron charge, $m$ is the charged
  particle mass and $c$ is the speed of light.
% where $r_c =e^2/(mc^2)$ is the ``electromagnetic radius'' of an
% electron ($e=4.803\times 10^{-10}\mbox{ e.s.u.}$ is the unit
% electric charge, $m$ is the electron mass and $c$ is the speed of
% light).  {\it This radius is assumed to be interpreted within the
%   framework of classical physics that all the mass of a charged
%   particle is brought about by the energy of static electric
%   field.})
%{\bf Not sufficiently accurate!!}

Clearly, Thomson cross section is much larger for electrons than
  for nuclei because $m_\text{e}\ll m_\text{p}$ (specifically for
  electrons, $\Sigma_\text{ph}=6.65\times 10^{-25}\mbox{ cm}^2$).
For a fully ionized plasma of mass density $\rho$, we obtain the
  mean free path of photons ($n_\text{nuc}$ and $n_\text{e}$ are
  the number densities of nuclei and electrons, respectively):
\begin{equation}
  \lambda_\text{ph}=\dfrac{1}{n_\text{nuc}
    \Sigma_\text{ph,nuc}+n_\text{e}\Sigma_\text{ph,e} }
    \simeq\dfrac{1}{\Sigma_\text{ph,e}n_\text{e}}\ .
\end{equation}
Numerically, we estimate a photon mean free path as
\begin{equation}
  \lambda_\text{ph}\simeq 5\mbox{ cm}\left(\dfrac{\rho}{1
      \mbox{ g cm}^{-3}}\right)^{-1}\ .
\end{equation}
This implies that photons are tightly trapped
  inside the cavity even for an envelope mass
  density as low as $\sim 10^{-5}\mbox{ g cm}^{-3}$.
% which is much smaller than $10^9\mbox{ g }\mbox{cm}^{-3}$
% for trapping neutrinos.
The much higher efficiency of momentum transfer (compared
  with neutrinos) also permits the possibility of
  radiation-driven envelope expansion of a void.

\subsection{Void solutions as asymptotic conditions}
\label{sec:ASYMPTOTIC}

\label{sec:DIMENSIONAL_MODEL}
\begin{figure}
    \includegraphics[bb=0 0 330 470, width=70mm]
    {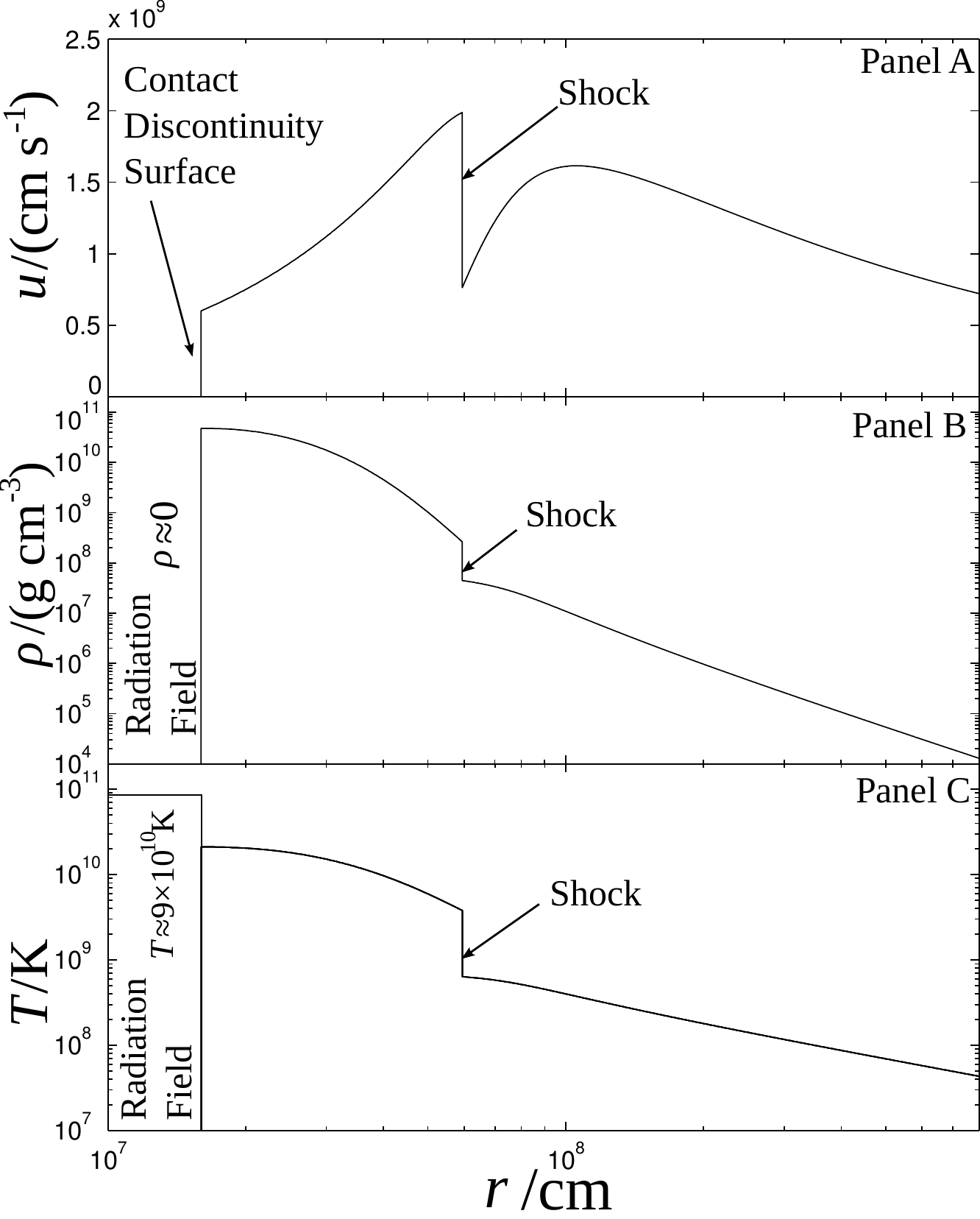}
    \caption{
%    {\bf Please use larger fonts.}
Dimensional model presentation at $t=t_i$ (the initial
    timescale of a self-similar void evolution).
Panels A, B and C illustrate the radial flow
      velocity (in unit of $10^9\mbox{ cm s}^{-1}$),
      mass density (in unit of $\mbox{g cm}^{-3}$) and
      temperature (in unit of Kelvin K) profiles of the
      model, respectively.
Vertical line-sections in the plot on the left represents the
      contact surface, inside of which is the intense radiation field,
      while those on the right indicates the expanding shock.
Note that we have also shown the temperature of the
      radiation field inside the central cavity in Panel C.}
%{\bf Please fully explore various plausible models
%     with physical parameters. Provide more astrophysical
%      contexts.
%      mass black hole formed during such explosion.
%An interesting example would be SN 1979C as reported
%in a recent paper.  }
    \label{fig:DIMENSIONAL_PROFILE_INITIAL}
\end{figure}

The mass is conserved in evolution and thus mass
  densities throughout the envelope decrease as
  time goes on for an overall expansion.
Since we investigate the situation that radiation field
  continues to drive the envelope outwards, when the
  gas in the inner envelope (especially near the
  contact discontinuity surface) becomes sufficiently
  tenuous, trapped photons can leak out gradually.
%our model mechanism is no longer valid.
%   will eventually fail in principle.
%
  % This time duration $\Delta t$ that our model analysis may be
  % applied is estimated with respect to the criterion that
  % $\rho_\text{cd}>\rho_c$, which has been discussed in
  % Sec. \ref{sec:PHOTON_SCATTERING}.

%{\it
We define $\Delta t$ as the time duration after $t_i$ (for
  its definition see Appendix C; hereafter we use subscript
  ``i" for those initial values in dynamic void evolution)
  to the time that the mass density near the contact
  discontinuity becomes fairly low (e.g.
  $\rho_\text{cd}\sim 10^{-5}\mbox{ g cm}^{-3}$,
  see subsection \ref{sec:PHOTON_SCATTERING}).

In reference to similarity transformation \eqref
  {eq:SELF_SIMILAR_TRANSFORMATION}, an estimate
  of $\Delta t$ is given by
%  }
\begin{equation}
  \Delta t\sim t_i\left[\left(\dfrac{\rho_{\text{cd},i}}
  {10^{-5}\mbox{ g cm}^{-3}}\right)^{1/2}-1\right]\ .
\end{equation}
% since $\rho\propto t^{-2}$ at constant $x$ and $\alpha$ (here
% $x_\text{cd}$ and $\alpha_\text{cd}$).  {\it This kind of upper
%   limit does not take diffusion effect near the contact
%   discontinuity into consideration, but diffusion is not
%   so severe as the decrease in mass density.}
%{\bf Need an explanation!!}
%{\it
Meanwhile, the diffusion across the contact
  discontinuity may modify our model.
In fact, diffusion process is not so effective
%  severe that we should take it into
%  significant consideration, which
  as analyzed in Appendix \ref{sec:DIFFUSION}.
%  }

% Although, in principle, our radiation-driven model cannot
%  survive the attenuation led by evolution, we can still
%  exploit the void solutions for descriptions on later stages.
%{\it
The mass density near the contact discontinuity is attenuated
  in a self-similar expansion, which would make our radiation
  driving expansion model invalid after a sufficiently long
  lapse in time.
We here show that our self-similar model may be
  further used as an asymptotic solution when the
  radiation-driving mechanism is no longer effective.
%  } {\bf Please clarify!}

Recalling self-similar transformation
  \eqref{eq:SELF_SIMILAR_TRANSFORMATION}, the pressure
  decreases with $p_\text{cd}\propto t^{2n-4}$, while the radius
  of the contact discontinuity expands with a power law $t^n$.
% Thus the ``force'' exerted upon the contact surface is proportional
%   to $t^{4(n-1)}$: if $n<1$, we can reasonably expect the force to
%   die out gradually.
 % {\it
We define the integral of radiation pressure
  $p_\text{cd}$ over the contact discontinuity
  surface as the ``radial radiation force" $F_r$,
\begin{equation}
  \label{eq:RADIAL_FORCE}
  F_r=4\pi r_\text{cd}^2p_\text{cd}=x_\text{cd}^2
  \alpha_\text{cd}^\gamma\dfrac{k^2t^{4n-4}}{G}\ .
\end{equation}
This result indicates that the force exerted on the gas shell by
  the radiation inside the bubble becomes less and less important
  as time goes on, since there is usually $4(n-1)<0$ (from
  $\gamma>1$ and $n+\gamma=2$).
For a long enough time $t$, the pressure force across the contact
  discontinuity surface diminishes to a negligible level.
In other words, even though there should be a certain pressure
  across the contact discontinuity surface, the dynamics of the
  shell are not significantly influenced if those pressures are
  weakened when $t$ becomes sufficiently large.
This analysis enables us to regard our void model to be valid as
  an asymptotic solution in the epoch when the gas shell becomes
  ``optically thin" and thus radiation in the bubble leaks out.
% }
%{\bf Clarification needed!}

As an example, for the model parameters specified in
  Table \ref {table:DIMENSIONAL_PROFILE_INITIAL}, if we take
  $\rho_\text{cd}\sim 10^{-5}\mbox{ g cm}^{-3}$ as the
  transition criterion from ``optically thick" to ``optically
  thin", we get
\begin{equation}
  p_\text{cd}|_{\rho_\text{cd}=10^{-5}\text{ g }\text{cm}^{-3}}
  =3\times 10^6\mbox{ dyn cm}^{-2}\ .
\end{equation}
This is a low pressure in comparison with
  the inertia of the stellar envelope.
We evaluate the ``radial radiation force"
  exerted by the radiation pressure as
\begin{equation}
  F_r=4\pi r_\text{cd}^2p_\text{cd}\sim 10^{33}\mbox{ dyn}\ .
\end{equation}
%While the total mass of the stellar envelope is $20M_\odot\sim
%  4\times 10^{34}\mbox{ g}$, the magnitude of acceleration is
%  estimated as $a\sim 10^{-2}\mbox{ cm s}^{-2}$.
%This leads to an extra displacement of $\sim a t^2\sim 10^8\mbox{
%    km}$, which is negligible when considering the radius of the
%  majority of ejecta: $\sim 10^{12}\mbox{ km}$ (if we set the typical
%  initial radius of the ejecta to be several times as $R_\odot$,
%  according to the conventional mass-radius relation in \citealt
%  {THEORETICAL_ASTROPHYSICS}).  {\bf Need to check! Implications!
%  Time $t$ value?}

% Meanwhile, the radial velocity of the void boundary scales
%   as $\propto t^{n-1}$.
% Therefore, energy consumption continues decreasing, rendering the
%   weight of central pressure less important (it is trivial to give
%   that the power of central force $\propto t^{5(n-1)}$).
% These offers more plausibility and legitimacy to finally
%   neglect the central radiation field in much later stages.

%{\bf Implication?}{\it I have commented this paragraph and stated
%    its essential meanings in the discussions related to eq. \eqref
%    {eq:RADIAL_FORCE}.}

As a result, even though central radiation field has gone and there
  may be many deviations from the original formation (e.g., the
  shock may eventually die out, asymmetry may become more apparent,
  diffusion across the contact surface may become severe), void
  solutions are still reasonable to the degree of (at least)
  approximations, since the inertia becomes predominant in
  further continuation of expansion.

\subsection{Self-similar evolution of SNe driven by\\
  \qquad\ central photon radiation pressure}

In this subsection, we briefly discuss a few specific
  applications of the conventional polytropic self-similar
  void solutions with shocks in the context of SNe.
The procedure of constructing a physical model from
  dimensionless solutions are summarized in Appendix C.

\subsubsection{A dynamic void model of self-similar evolution}
\label{sec:DYNAMIC_EXAMPLE}

\begin{table}
\centering
\caption{
Parameters specifying conventional polytropic void models
  with shocks; in dimensional form, they describe the
  dynamic evolution of a gas shell, e.g. a SN.
``CD" in this Table denotes the contact discontinuity. ``Duration
  time" is the time span of the model validity considering the
  scattering process of the charged particles by
  photons (see subsection \ref{sec:ASYMPTOTIC} for details).}
    \begin{tabular}{ccc}
      \\
      \hline
      Item  & Variable Name  & Value \\
      \hline
      Total mass & $M$ & $\sim 20M_\odot$ \\
      Density at CD    & $\rho_{\text{cd},i}$ &
      $\sim 4.7\times 10^{10}\mbox{ g }\mbox{cm}^{-3}$\\
      Pressure at CD   & $p_{\text{cd},i}
      %{\bf Check unit!!}
      $ & $\sim
      10^{29}\mbox{ dyn }\mbox{cm}^{-2}$\\
      Initial void radius& $r_{\text{cd},i}$ &$\sim 160\mbox{ km}$ \\
    Radial velocity of CD& $u_{\text{cd},i}$ &$\sim 6000\mbox{ km s}^{-1}$ \\
      Cavity Temperature & $T_{\text{rad},i}$&$\sim 9\times 10^{10}\mbox{ K}$\\
      Duration time      & $\Delta t$        &$\gsim 10^6 \mbox{ s}$ \\
      \hline
    \end{tabular}
\label{table:DIMENSIONAL_PROFILE_INITIAL}
\end{table}

By properly specifying similarity transformation and envelope
  cutoff (see Appendix C), we would adopt our Model 6 with key
  parameters summarized in Table \ref{table:PARAMETER_WITH_SHOCK}
  to describe a self-similar void evolution for a model SN.
The progenitor has a mass of $\sim 20M_\odot$.
 An envelope cutoff radius is set at
  $r_\text{cut}\simeq 40R_\odot$ where the mass density is
  $\rho_\text{cut}\simeq 10^{-8}\mbox{ g cm}^{-3}$
  and the gas temperature is taken as $T\simeq 4000\mbox{ K}$.
%  {\bf Please clarify!}

Before the phase of self-similar evolution, a cavity around
  the center has already been carved out by the powerful
  neutrinosphere, before neutrinos decouple from
  the surrounding envelope and escape.
The cavity radius is initially estimated as $\gsim 100\mbox{ km}$,
  by Bethe \& Wilson (1985), Bethe (1990)
%  \citet{1985ApJ...295...14B, RevModPhys.62.801}
  and later works on the Wilson mechanism such as
  \citet{1996A&A...306..167J}.
Here we take the initial cavity radius $r_{\text{cd},i}$
  to be $\sim 160\mbox{ km}$.
% {\bf Why is this difference?}
Subsequently, the stellar envelope surrounding
  the cavity expands outwards
%  with its own inertia,
  continuously powered by photon radiation pressure
  during the following self-similar evolution.
Some important numerical values of parameters for
  describing this model (which also specifies
  the similarity transformation) are given in
  Table \ref{table:DIMENSIONAL_PROFILE_INITIAL}.

% {\it In Table 2, $\Delta t$ is worth noting. It indicates
%   the time period during which our model remains valid.
%   This validity is considered with respect of the efficiency of
%   momentum transfer from the radiation field to the gas (see
%   Appendix C for details).  However, this $\Delta t$ is just a
%   reference. First, there is another factor undermining the
%   validity of this model: the diffusion effect.  In spite of this,
%   since the diffusion is not severe (around $1.3\%$ as the radius
%   of central void doubles), this $\Delta t$ is still a fair
%   indication of the time scale for the model validity.  Secondly,
%   even though our model seems to become invalid as a result of
%   attenuation, it could still be a reasonable (although rough)
%   approximation of similarity evolution, which we are going to
%   discuss later.}
%{\it
While $\Delta t$ (defined in subsection
  \ref{sec:ASYMPTOTIC}) is presented in Table
  \ref{table:DIMENSIONAL_PROFILE_INITIAL}, we note that the
  model can remain valid after this $\Delta t$ as a asymptotic
  condition noted in subsection \ref{sec:ASYMPTOTIC}.
%}  {\bf Clarification needed!!}

As an example, radial flow velocity, density and temperature
  profiles at the initial stage ($t=t_i$) are displayed in
  Fig. \ref{fig:DIMENSIONAL_PROFILE_INITIAL}.
%
% The shock is not the rebound shock that galvanizes the explosion:
%  indeed, the essentials of the radial speed profiles have already
%  been formed by the acceleration of the rebound shock.
%{\bf What is the point?}{\it I have searched several articles which
%  tells me that it will be possible to see this shock as the
%  rebound shock. So I commented this statement.}
%
From Panel A of Fig. \ref{fig:DIMENSIONAL_PROFILE_INITIAL}, we
  observe the impact of the radiation-driven mechanism in the
  dynamic evolution of envelope with an expanding central void.
While the radial flow speed throughout the shell remains
  always positive and large, there are still intervals of
  radius where gas materials show an impressive trend to
  collapse inwards under self-gravity.
In fact, a quick glance at some other self-similar models with
  $\gamma$ in EoS being close to $4/3$ suggests a strong trend
  of the shell to decelerate (or even collapse) in the upstream
  side of a shock (see Models 5 and 6).
The expanding shock here serves as an important and effective
  accelerator to drive the envelope outwards against self-gravity.
% However, shock propagation is a dramatic energy consumer. In other
%  words, if the expansion relies solely on the envelope inertia,
%  it would be possible for the expanding envelope to decelerate much
%  faster than observations expect, or even begin to collapse.
%{\it However, shock propagation needs a lot of energy.
%If the radiation field inside the bubble is absent in the early epoch of
%  expansion (in late epoch we could find this absence is not dramatic,
%  see the discussions in Sec. \ref{sec:ASYMPTOTIC}), the shock will
%  possibly die out. }
%{\bf Explanation!}

Regarding our analysis of self-similar model, the photon radiation
  field trapped inside the central bubble appears to be a physically
  plausible candidate of the energy source, which drives the
  envelope outwards continuously during the optically thick phase.
From Panels B and C of Fig. \ref{fig:DIMENSIONAL_PROFILE_INITIAL},
  we readily recognize an outgoing shock.
It is a strong shock with a drastic jump across the
  shock front in both mass density and temperature.
The contact discontinuity surface at
  $r_{\text{cd},i}=160\mbox{ km}$ initially, on the
  other hand, was supported by the strong neutrino
  flux before the decoupling of neutrinos.
The neutrino flux also heats up rarified materials inside the
  void significantly and is responsible for the discontinuity
  in temperature across the contact interface.

\subsubsection{Supernova SN1993J}
\label{sec:SN1993J}

\begin{figure}
  \includegraphics[bb=0 0 330 460, width=70mm]{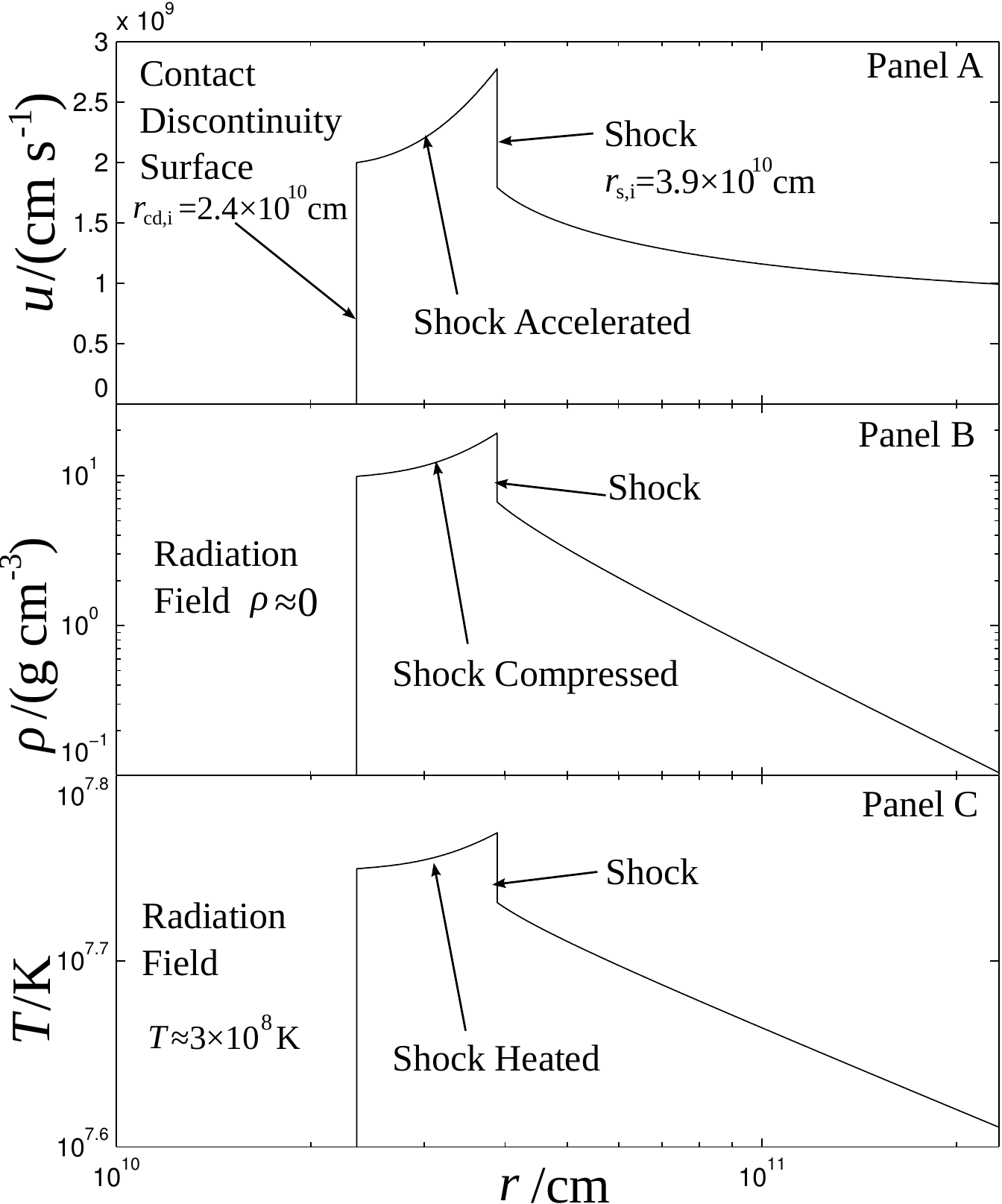}
  \caption{
Possible fittings of radial velocity ($u$ in Panel A), mass
  density ($\rho$ in a logarithmic scale in Panel B) and
  temperature ($T$ in a logarithmic scale in Panel C) versus
  radius ($r$ in a logarithmic scale) profiles of SN1993J ejecta
  in the early stage (a timescale of $\sim 10\mbox{ s}$).
The ``shell" region between the contact discontinuity and the
  shock is $\sim 30\%$ of the shock radius, in accordance with
  radio observations \citep[e.g.][]{2009A&A...505..927M}. }
  \label{fig:SN1993J_INITIAL}
\end{figure}

SN1993J in M81 is a type-IIL core collapse supernova about
  $3\mbox{ Mpc}$ away from us \citep[e.g.][]{1993Natur.364..600S}.
This close neighborhood has enabled a plethora of observations.
Here we take the observation and simulation results presented
  in \citet{2011A&A...526A.142M, 2011A&A...526A.143M} and
  references therein.
SN1993J has a spherical asymmetry $\lsim 2\%$ and it
  would be of considerable interest to explore the
  capability of our void model with spherical symmetry.
Numerical simulations of
  \citet{2011A&A...526A.142M,2011A&A...526A.143M} have
  adopted the dynamic model of \citet{1982ApJ...258..790C}
  where the self-gravity is neglected.

There are different fittings of SN1993J expansion presented
  in \citet{2011A&A...526A.142M} with various expansion
  parameter $m$ (not the reduced mass here)
%; definition is similar {\bf Meaning?}
%  to \citealt{2011A&A...526A.142M})
  which takes the similar connotation as $n$ in our formulation.
%{\it
An accepted selection of $n$ seems to be $n=0.933$ before
  $\sim 360\mbox{ days}$ [similar to the selection specified
  in \citet{2009A&A...505..927M}].
Here, $t\sim 360\mbox{ days}$ is observed as the ``break" time,
   after which an observable deceleration in expansion occurs
%}
%Definition and meanings of this $n$ has already been discussed in
  (see subsection \ref{sec:EOS} for the implication of $n$).
%{\bf Clarify!}

Some model features are  noted here.
%{\it
Several simulations such as \citet{2011A&A...526A.142M} have
  suggested that the radiation opacity of the shell has been
  fitted to be more than $\sim 80\%$ during
  $1\mbox{ yr}\sim 3\times 10^7\mbox{ s}$.
This implies that our model may be valid
  during the first $\sim 360\mbox{ days}$.
%}
This justifies our void model to be applied even in later times
  since the pressure is actually low enough to be ignored
  dynamically (see subsection \ref{sec:ASYMPTOTIC}).

Here we briefly describe a possible version of dimensional
  dynamic model for certain features of SN1993C ejecta.
The mass of the progenitor is $\sim17M_\odot$
  \citep[e.g.][]{1994AJ....107..662A}.
The radio bright shell consists of $\sim 30\%$ of the radius \citep
  [e.g.][]{2009A&A...505..927M} (it is taken that this radio bright
  region consists mainly of the shocked gas between the shock and
  the contact discontinuity) and so forth.

For our void model, the mass cutoff is set at
  $r_\text{cut}\simeq 6R_\odot$, where the mass density is
  $\rho_\text{cut}\sim 0.03\mbox{ g cm}^{-3}$, according
  to the mass-radius relation in \citet{THEORETICAL_ASTROPHYSICS}.
% This radius is not severely impacted at such a time scale
% when the ``surface" takes an outward displacement {\it less}
% than $\sim 10^5\mbox{ km}\ll R_\odot$ during the .

The ``bright" region is assumed to be between the shock surface
  and the contact discontinuity where the gas is compressed and
  shock heated and charged particles (electrons in particular)
  are dramatically accelerated across the magnetized shock front
  -- this gives considerable radio emissions which can be
  detected by radio telescopes.
Expansion of the shocked spherical shell is self-similar;
  specifically, the shock radius $r_\text{s}$ will expand by
\begin{equation}
  r_\text{s}=3.9\times 10^{10}\mbox{ cm}
  \left(\dfrac{t}{t_i}\right)^{0.933},
\end{equation}
where the timescale $t_i$ is fitted to be $11\mbox{ s}$.
 If we input $t$ to be $330\mbox{ days}$ after the SN
 explosion, we would expect the radius of the shock
 front (hence the radius of detectable radio emissions)
 to be $r_\text{s}=3.8\times 10^{16}\mbox{ cm}$.
Taking SN1993J to be $3\mbox{ Mpc}$ away
  and neglecting cosmological effects
% [$3\mbox{ Mpc}$ is far from important in cosmology, e.g. see
  [e.g., \citet{RYDEN_COSMOLOGY}], we would then
  have the angular radius $\delta\theta$ as
\begin{equation}
  \delta\theta\simeq\dfrac{r_\text{s}}{3\mbox{ Mpc}}
  \simeq 0.85\mbox{ mas}\ ,
\end{equation}
  which is in very good agreement with the result of VLBI
  observations [$\sim (0.818\pm 0.015)\mbox{ mas}$]
  presented in \citet{2009A&A...505..927M}.
Fig. \ref{fig:SN1993J_INITIAL} shows these results for comparison.

\section{Summary and Conclusions}
\label{sec:CONCLUSION}

We have systematically explored various conventional
 polytropic (i.e. $n+\gamma=2$) self-similar void
 solutions by adopting self-similarity transformation
 (\ref{eq:SELF_SIMILAR_TRANSFORMATION}) in nonlinear
 Euler hydrodynamic PDEs of spherical symmetry.
Different types of dynamic void solutions with outgoing
 shocks are constructed numerically, while those with
 $n=2/3+\epsilon$ and $\epsilon>0$ (e.g., $n=0.67$)
 are described in more details.
While highly idealized, these are dynamic void solutions
 with expanding shocks into various possible envelope
 types very close to the so-called ``hot bubble" or
 cavity cases with potential astrophysical counterparts.
Combined with proper central sources of energy and
 momentum, such void solutions can be initiated and
 sustained by matching the pressure balance condition
 across the contact discontinuity in expansion during
 a certain phase of self-similar evolution.
We discussed several possible situations.

For supernova explosions of massive progenitors, we advance
 the following physical scenario for the gross separation
 between a more massive envelope and a less massive
 collapsing central compact object.
Shortly after the onset of core collapse of a massive
 progenitor star, the powerful neutrinosphere trapped
 in an extremely dense core drives outwards to carve
 out a rarified zone of one or two hundred kilometers.
As energetic neutrinos escape from the stellar envelope
 of decreasing density, intense photon radiation field
 and/or electron-positron pair plasma mixed together
 continue to drive the central cavity outwards against
 the envelope.
When the mass density of the expanding envelope becomes
 sufficiently low, photons leak out effectively in the
 optical thin regime eventually.
If the pressure across the contact discontinuity interface
 diminishes to a sufficiently low level or diffusion effects
 (see Appendix D) smear out the ``contact discontinuity" at
 sufficiently large radii, the outer remnant envelope
 continues to expand by inertia into the tenuous
 interstellar medium.

Specifically for our model construction, we have
 first explored possible conventional polytropic
 void solutions crossing the SCC smoothly.
Similar to those isothermal voids shown by
 \citet[][2010]{springerlink:10.1007/s10509-009-0044-4}, there
 are several intervals of the independent similarity location
 of the sonic critical point $x_0$ --- some of which permit the
 existence of such type of void solutions, while others do not.
The SCC are divided into two segments for $n\neq 1$, and
 there are no such void solutions crossing SCCs at type-2
 critical points in segment 1 of SCC for $n<0.84$.
In fact, there is no type-2 void solution with $x_0<50$
 for $n$ less than $0.84$, as we have found in extensive
 numerical explorations.

Self-similar void solutions with outgoing shocks are also
 constructed, with several different kinds of envelopes outside
 the expanding shock, including the static SPS envelope, the
 outflow, inflow, breeze and contraction envelopes.
Through extensive numerical explorations, we found no void
 solutions with shocks propagating into a static SPS, or
 breeze, or contraction envelope for $n$ less than $0.80$.
%
% have figured
% out that there will be no EWCS-envelope
% (or SPS-envelope, as special cases) shock solutions with
% voids for $n$ less than $0.80$, and thus breeze and
%  contraction solutions will also disappear.
 % For an $n$ value close to $2/3$, our model predicts a dramatic
 % deceleration of envelope expansion led by self-gravity, which we
 % have discussed about in deceleration led by self-gravity.
% {\bf Clarification!}{\it It is so obvious from the similarity
%   transformation that the deceleration is severe when $n$ is
%   considerably smaller than $1$ that I deleted the original
%   statement.}

We have constructed some dynamic void solution examples,
 especially in the context of SNe, as applications of
 our self-similar void models.
For hot bubbles, $\gamma$ values are usually $4/3$ (see Lou \& Cao
 2008 for a general polytropic gas)
% \citep[see][for a general polytropic gas]{2008MNRAS.384..611L}
 and for conventional polytropic gas dynamics,
 the corresponding $n$ should be $2/3$ exactly.
In reality, we expect deviations from the exact $n=2/3$ cases.
 We have thus discussed deviations of $n$ values from $2/3$,
 especially the physical reality of $\epsilon$ in
 $n=2/3+\epsilon$.
%  : for an $\epsilon$ less than $1/15$,
% $2/15$,
% it represents a certain kind of energy
% input in the central void.
In addition, we have discussed the diffusion effect at void
 boundaries in Appendix D, finding that the diffusion is
 around merely $1.3$ percent as the central cavity radius
 nearly doubles.
We have compared the roles of radiation field and neutrino
 flux respectively after the neutrinosphere having
 decoupled from the gas shell with the result that radiation
 field would be a dominant force in the evolution of SN ejecta
 while the ejecta tend to be transparent to neutrinos.
Also in subsection \ref{sec:ASYMPTOTIC}, we have discussed
 applications of our model as an asymptotic condition when
 the pressure across the contact discontinuity diminishes
 to insignificant level.

Finally, we have attempted to specify our model parameters
 to describe dynamic evolution of astrophysical objects
 and construct dimensional profiles in Fig.
 \ref{fig:DIMENSIONAL_PROFILE_INITIAL}.
%As a trial, assuming that our model is still valid when the
% central radiation field has become weaker (justification of this
% assumption has been discussed), we have raised a possible version
% of the expansion profiles of SN1979C ejecta, {\bf Still want this
% case?!} which implies it
An example of application is the SN1993J. The dynamic model
 has been used to portray SN1993J evolution of expansion before
 the ``break time" $t_\text{br}$, where we have found that the
 results are able to capture some characteristics presented by
 radio observations, such as a $\sim 30\%$ radius bright shell,
 the angular radius and the total stellar mass.

\section*{Acknowledgments}
This research was supported in part by Tsinghua Centre for
  Astrophysics (THCA), by the National Natural Science
  Foundation of China (NSFC) grants 10373009, 10533020,
  11073014 and J0630317 at Tsinghua University, by Ministry
  of Science and Technology (MOST) grant 2012CB821800,
%  射电波段的前沿天体物理课题及FAST早期科学研究 公示内容
  and by the Yangtze Endowment, the SRFDP 20050003088 and
  200800030071, and the Special Endowment for Tsinghua
  College Talent Program from the Ministry of Education
  at Tsinghua University.
%This research has been supported in part by the ASCI Centre for
%Astrophysical Thermonuclear Flashes at the University of Chicago,
%under the Department of Energy contract B341495,
%by the Special Funds for Major State Basic Science Research
%Projects of China, by the Tsinghua Centre for Astrophysics,
%by the Collaborative Research Fund from the National Science
%Foundation of China (NSFC) for Young Outstanding Overseas
%Chinese Scholars (NSFC 10028306) at the National Astronomical
%Observatories, Chinese Academy of Sciences,
%by the NSFC grants 10373009 and 10533020 at the
%Tsinghua University, and by the SRFDP 20050003088
%Specialized Research Fund for the Doctoral Program of Higher
%Education and the Yangtze Endowment from the Ministry of
%Education at the Tsinghua University.
%Affiliated institutions of Y-QL share this contribution.

\appendix

\section{Properties for the jump of $\alpha$
  from zero at the zero mass line }\label{sec:APPENDIX}

At the zero mass line (ZML) where $nx=v$, eq. \eqref{eq:DYNAMIC_ODES}
 shows a singularity for $\mbox{d}v/\mbox{d}x$ and $\mbox{d}\alpha
 /\mbox{d}x$ if we simply take $\alpha=0$ there with $\gamma>1$.
An alternative approach is to let $\alpha\rightarrow 0$, which
 leads to
\begin{equation}
    \left. \dfrac{\mbox{d}\alpha}{\mbox{d}x}\right|_{nx=v}
    \rightarrow 0\ ,\qquad\qquad\left. \dfrac{\mbox{d}v}{\mbox{d}x}
    \right|_{nx=v}\rightarrow 2(1-n)\ ,
\end{equation}
according to eq. \eqref{eq:DYNAMIC_ODES}.\footnote{We can
  also attain these results by setting $\alpha=0$ in related
  equations (16) and (17) of \cite{Lou2010198}.}
%  {\bf Sure of this?} {\it
The second derivative of $v$ with respect
 to $x$ can be calculated below as
%\begin{equation}
%\begin{split}
%    \left. \dfrac{\mbox{d}^2 v}{\mbox{d}x^2}\right|_{nx=v}=
%    & \dfrac{1}{\gamma\alpha^{\gamma -1}}\bigg[2\gamma
%    \alpha^{\gamma-1}\dfrac{2n-1+(n-1)x}{x}
%    \\
%    + & 2\gamma (\gamma-1)\alpha^{\gamma-2}(1-n)
%    + nx(1-n)(3n-2)\bigg]
%    \\
%    - & \dfrac{2(1-n)(\gamma-1)}{\alpha}\ .
%\end{split}
%\end{equation}
%It is readily seen that $\mbox{d}^2v/\mbox{d}x^2$ vanishes
% in an isothermal gas with $n=\gamma=1$. {\bf Not obvious!}
%Nevertheless with $\gamma\neq 1$ (thus $n\neq 1$ in a conventional
% polytropic gas), we get $\mbox{d}^2v /\mbox{d}x^2\rightarrow
%\infty$ for $\alpha\rightarrow 0$. Therefore, we cannot obtain a
%physically acceptable solution for the condition that $\alpha=0$
%at ZML in the conventional polytropic model of self-similar
%evolution.
%***************************************************
%The following is added later as the commented out
%paragraph above.  August 21, 2011
\begin{equation}
  \left. \dfrac{\mbox{d}^2 v}{\mbox{d}x^2}\right|_{nx=v}=
  \dfrac{(3n-2)}{x}\left[2+\dfrac{n(n-1)x}{(n-2)}\alpha^{n-1}
  \right]\ .
\end{equation}
It is readily seen that $\mbox{d}^2v/\mbox{d}x^2$ is
  equal to $2/x$ in an isothermal gas with $n=\gamma=1$
  \citep[see eq. 23 of][]{Lou2010198}.
Nevertheless with $\gamma>1$ (thus $n<1$ in a
  conventional polytropic gas), we would have
  $\mbox{d}^2v/\mbox{d}x^2\rightarrow \infty$
  for $\alpha\rightarrow 0$.
Moreover, this $\alpha|_{nx=v}\rightarrow 0$ would also give
  a solution with zero $\alpha$ everywhere (as a function of
  $x$), which has also been verified by \citet{Lou2010198}.
Therefore, we cannot obtain a physically acceptable
  solution for $\alpha=0$ at ZML in the conventional
  polytropic void model of self-similar evolution.

\section{Thermodynamic Derivations of Energy Input Rate}

In Appendix B here, all quantities are for
  the radiation field inside a bubble.
From the energy conservation, we have
\begin{equation}
  \mbox{d}U=\mbox{d}Q-p\mbox{d}V\ ,
\end{equation}
where $\mbox{d}Q$ is a differential
  heat transfer into a bubble.
Radiation and relativistic matter have the internal
  energy $U=3pV$ where $p$ is the pressure and $V$
  is the bubble volume, and thus
\begin{equation}
  \mbox{d}Q=4p\mbox{d}V+3V\mbox{d}p\ .
\end{equation}
If the EoS takes the form of $p\propto
  \rho^{4/3-\epsilon}$, we would have
\begin{equation}
  \dfrac{\mbox{d}p}{p}=\left(\dfrac{4}{3}-\epsilon\right)
  \dfrac{\mbox{d}\rho}{\rho}\ =-\left(\dfrac{4}{3}-\epsilon\right)
  \dfrac{\mbox{d}V}{V}\,
\end{equation}
and it follows immediately that
\begin{equation}
  \mbox{d}Q=3\epsilon p\mbox{d}V\ .
\end{equation}

\section{From Dimensionless Solutions to Physical Models}

%{\it
In Appendix C here, we discuss how to produce
  a physical model from the reduced dimensionless
  solutions of ODE \eqref{eq:DYNAMIC_ODES}.
It is important to reasonably specify the self-similar
  transformation (there are some DOFs) and hence
  consistently interpret one dimensionless solution
  for a physical model.
We can then determine the subsequent self-similar
  evolution of a void surrounded by a shocked envelope.

A convenient means to consistently specify the transformation
  parameters is to relate the initial values of physical
  quantities of the flow system on the contact discontinuity
  (e.g., the mass density $\rho_\text{cd}$) to the corresponding
  dimensionless variables (e.g. $\alpha_\text{cd}$) with respect
  to eq. \eqref{eq:SELF_SIMILAR_TRANSFORMATION}.
%  }
We shall attach a subscript ``$i$" for those initial values
  of physical variables in a self-similar hydrodynamic evolution.

First we consider the time $t$. With self-similar models adopted
  here, the time $t$ cannot be zero even at the very first stage
  of similarity evolution, because $t=0$ is either a zero point
  ($n\geq 1$) or a singular point ($0<n<1$) of the self-similar
  transformation \eqref{eq:SELF_SIMILAR_TRANSFORMATION} and
  resulting ODE \eqref{eq:DYNAMIC_ODES}.
%{\it
A value $t_i$, indicating the value of time
  $t$ at the initial phase of self-similar evolution,
  needs to be specified or chosen.
The self-similar transformation
  \eqref{eq:SELF_SIMILAR_TRANSFORMATION}, from
  the initial mass density at the contact discontinuity
  surface $\rho_{\text{cd},i}$ to $\alpha_\text{cd}$,
  enables us to estimate $t_i$ as
% }{\bf Please explain!}
\begin{equation}
\label{eq:INITIAL_TIME}
  t_i=\left(\dfrac{\alpha_\text{cd}}{4\pi G
      \rho_{\text{cd},i} }\right)^{1/2}
  = 0.011\mbox{ s}\ \alpha_\text{cd}^{1/2}
  \left(\dfrac{\rho_{\text{cd},i}}{10^{10}
      \mbox{g cm}^{-3}}\right)^{-1/2}.
\end{equation}
A similar $t_i$ is introduced as the ``cut-off time"
  in \citet{2006MNRAS.372..885L} with respect of
  shock evolution; alternatively, we here focus
  on various dynamic evolution of voids.

%Since the sound parameter $k$ in self-similar transformation
%  \eqref{eq:SELF_SIMILAR_TRANSFORMATION} is independent of $t$,
We can also determine the value of sound parameter $k$ at $t_i$
  which is still valid in the subsequent evolution (for the
  presence of a shock in self-similar evolution, this $k$
  would then be $k_2$ for the downstream flow)
\begin{equation}
  k^{1/2} =r_{\text{cd},i}x_\text{cd}^{-1}t_i^{-n}\ .
\end{equation}
% Here and above in eq.\eqref{eq:INITIAL_TIME}, the presence of
% dimensionless variables, $x_\text{cd}$ and $\alpha_\text{cd}$,
% reflects the dependency of the overall scenario of evolution on
% the self-similar model (dimensionless model) chosen.
%
%{\bf Meaning?}{\it I have stated the essential meanings of those
%  confusing statements in the beginning of this appendix, which
%  are in italic.}
%
% Thus the necessity of choosing a compatible model with
%  respect of the real-world profiles in density, radial velocity,
%  etc., is obvious.
% In other words, those variables imply that we have some degrees of
%  freedom in choosing model, but the selection must be based on
%  physical reality.
% {\bf Need a clarification!}
%
Another indispensable consideration deals with the
  cutoff enclosed mass and the cutoff mass density.
For
% a hydrodynamic system described by
  our self-similar void
  model, we need to choose a sensible enclosed mass $M$.
For a very large $r$ corresponding to $x\gg 1$
  at a fixed $t$, asymptotic solution
  $\alpha\propto x^{-2/n}$
  (eq. \eqref{eq:ASYMPTOTIC_SOLUTIONS}) yields
\begin{equation}
  m=\alpha x^2 (nx-v)\sim n x^{3-2/n}\ .
\end{equation}
As $x\rightarrow\infty$, $m\rightarrow\infty$
  for $n>2/3$ and $m\rightarrow 0$ for $n<2/3$.
%  {\bf Statement reversed?}
Neither of them is physically acceptable as $M$
  is proportional to $m$ and $M/m$ is independent
  of radius in our self-similar solution.
In dealing with this problem it is sensible to
  set a cutoff radius within which materials are
  included in the stellar mass;
beyond this cut-off radius, materials are not
  regarded as stellar mass.
% Legitimacy of adopting this cutoff has much to do with density at the
% cutoff radius: the mass density there should be low enough.
%{\it
In order to check whether the cutoff is proper, we can introduce
  the ``cutoff density" -- the mass density at the cutoff radius,
  as a criterion, which should be low enough for a reasonable cutoff.
%  } {\bf Please be precise!!}
The cutoff radius is usually large but constant.

With the help of asymptotic solution
  \eqref{eq:ASYMPTOTIC_SOLUTIONS}, we derive the
  following expressions for the enclosed mass at the
  cutoff radius $r_\text{cut}$ which is large enough:
\begin{equation}
  \label{eq:CUTTING_MASS}
  \begin{split}
    \dfrac{M_\text{cut}}{M_\odot}&=2\times 10^{10}(1.44\times
    10^{-5})^{2/n}\dfrac{Anx_\text{cd}^{-2/n}}{(3n-2)\alpha_\text{cd}}
    \eta_k^{1/n-3/2}
    \\
    & \!\!\!\!\!\!
    \times\left(\dfrac{r_\text{cut}}{R_\odot}\right)^{3-2/n}
      \left(\dfrac{r_{\text{cd},i}}{10^6\mbox{ cm}}
    \right)^{2/n}\left(\dfrac{\rho_{\text{cd},i}}{10^{10}
        \mbox{ g cm}^{-3}}\right),
\end{split}
\end{equation}
%{\bf Need to verify!}
and the cutoff mass density
\begin{equation}
\label{eq:CUTTING_DENSITY}
\begin{split}
  \dfrac{\rho_\text{cut}}{10^9\mbox{g }\mbox{cm}^{-3}}
  & =0.99\ (1.44\times 10^{-5})^{2/n}A
  \eta_k^{1/n}x_\text{cd}^{-2/n}\alpha_\text{cd}^{-1}
  \\
  & \!\!\!\!\!\!\!\!\!\!\!\!
  \times\left(\dfrac{r_\text{cut}}{R_\odot}\right)^{-2/n}
  \left(\dfrac{r_{\text{cd},i}}{10^6\mbox{ cm}}
  \right)^{2/n}\left(\dfrac{\rho_{\text{cd},i}}{10^{10}\mbox{g
      }\mbox{cm}^{-3}}\right).
\end{split}
\end{equation}
Notations for $A$, $n$ and $k$ in these expressions
 carry the exactly same definitions as those in
 eq. \eqref{eq:SELF_SIMILAR_TRANSFORMATION}.
Still, the sound parameter $k$ here is for the
 downstream side, $k_2$, in the presence of a shock.
%What seems to be a little strange is that the enclosed
% mass at a large radius is independent of time.
%  {\bf Need a discussion!}
%{\it
We demonstrate below that the enclosed
 mass at a large $x$ is independent of time $t$.
At a large radius $r$ for a given time
 $t$, the enclosed mass is evaluated by
\begin{equation}
  \begin{split}
    M\propto\ & t^{3n-2}m =t^{3n-2}\alpha x^2(nx-v)
    \\
    \simeq\ & nt^{3n-2} Ax^{3-2/n}\propto t^{3n-2}
    (t^{-n})^{3-2/n}\propto t^0\ .
\end{split}
\end{equation}
In the above derivation, we neglect $v$ in $(nx-v)$ because
  of a small $v$ as compared with $x$ in reference to eq.
  \eqref{eq:ASYMPTOTIC_SOLUTIONS}.
The leading order term of $x$ in
  eq. \eqref{eq:ASYMPTOTIC_SOLUTIONS} says, when $x\gg 1$,
  there is $v\sim x^{1-1/n}$.
Therefore $v/x\sim x^{-1/n}$ and converges to
  zero when $x\rightarrow\infty$ (for $n>0$).
This independence of time can also be explicitly seen from the
  fact that the radial flow velocity is usually small at large radii.
Thus, the radial flow of materials at large $r$ can be neglected
  in terms of its effect on the total mass enclosed by the spherical
  surface there.
%  }
% {\it We can see from eqs. \eqref{eq:CUTTING_MASS} and
%  \eqref{eq:CUTTING_DENSITY} that} for a value of $n$ around $2/3$,
%  $M_\text{c}$ depends on $r_c$ very insensitively; but $\rho_\text{c}$
%  is determined by $r_c$ very sensitively.
% That is in accordance with our general picture constructed for a
%  massive progenitor star \cite[e.g.][] {THEORETICAL_ASTROPHYSICS}.

\section{Diffusion across the Contact Discontinuity Interface}
\label{sec:DIFFUSION}

The balance of pressures across a contact discontinuity
 is a necessary condition for the establishment of the
 discontinuity interface bounding an expanding void during
 the process of self-similar hydrodynamic evolution.
Yet a dramatic difference of mass concentration (i.e.
 difference in chemical potential, to be precise)
 should lead to diffusion across the contact
 discontinuity interface.
%and thus undermine the applicability
%of the model discussed in this article.
We now estimate diffusion effects across the contact
 discontinuity surface surrounding the central void --
 more in the context of a supernova explosion.
Somewhat different from what has been explored by
 \citet[][2010]{springerlink:10.1007/s10509-009-0044-4}, the
 envelopes or gas shells are at a high temperature range
 of $\sim 10^{10}-10^{11}\mbox{ K}$ that relativistic
 kinetic effects for electrons cannot be ignored.
% neglected trivially, especially to those electrons.
Electrons diffuse faster than protons (or other nuclei)
 do as they are much lighter in mass, but collectively
 they are tightly trapped by the electric field produced
 by the difference of motions between electrons and
 positively charged nuclei.
% {\bf Need a clarification!}

Matters, including electrons, neutrons, nuclei,
  are almost completely degenerate inside the
  highly condensed core prior to the onset
  of a stellar core collapse.
However, we conclude from the following calculations
  [see
%, which is mostly in reference of
  \cite{CALLEN_THERMODYNAMICS}] that those materials
% {\bf This may not be true!}
  in the gas shell shortly after a SN explosion
  are far from degenerate.
The criterion of non-degeneracy is given by ($\mu$ is
  the chemical potential and $\beta=(k_{\text{B}}T)^{-1}$ with
  $k_{\text{B}}=1.3807\times 10^{-16}\mbox{ erg K}^{-1}$
  being the Boltzmann constant)
\begin{equation}
  e^{\beta\mu}\ll 1,
\end{equation}
%{\bf You mean $\beta^{-1}$?}
or equivalently,
%page 402
\begin{equation}
  \dfrac{\lambda_T^3{\cal N}}{g_0}\ll 1\ ,
\end{equation}
(see \citealt{CALLEN_THERMODYNAMICS}) where $\lambda_T
  =\left(2\pi mk_{\text{B}}T/h^2\right)^{-1/2}$ (here $h =
  6.626\times 10^{-27}\mbox{ erg }\mbox{s}$ is the Planck
  constant) is the thermal de Broglie wavelength, ${\cal N}$
  is the number density of particles, $m$ is the particle mass,
%{\bf Not to be confused with the scaling index $n$!!}
 and $g_0$ is the intrinsic degree of degeneracy;
 e.g., for particles with a spin quantum number
 $s$, we have $g_0=2s+1$.
% {\bf Pay attention to chemical potential! }
Thus we can derive the degeneracy condition of an electron
 gas (note that ${\cal N}$ can be estimated by
 ${\cal N}\simeq\rho/(m_\text{p}{\cal A})$, where $m_\text{p}$
 is the proton mass and ${\cal A}$ is the average nucleon
 number per nucleus), viz.
\begin{equation}
  \rho\gsim 2m_{\text{p}}\left(\dfrac{2\pi m_{\text{e}}
  k_{\text{B}}T}{h^2}\right)^{3/2}\simeq 10^{13}\mbox{ g cm}^{-3}
  \left(\dfrac{T}{10^{10}\mbox{ K}}\right)^{3/2}\ ,
\end{equation}
%{\bf Define notations! Give references!}
 where $m_{\text{e}}$ is the electron mass.
Here we assume that the numbers of protons and neutrons
 in nuclei in the gas are roughly the same.
This shows that materials in the core with as high mass
 density as $\rho\sim 10^{14}\mbox{ g }\mbox{cm}^{-3}$
 \citep[e.g.][]{1977ApJ...218..815A} are strongly degenerate.
Meanwhile, this justifies the adoption of the
 Maxwell-Boltzmann (M-B) statistics (i.e. non-degenerate)
 for electrons at mass density $\rho\lsim 10^{10}\mbox{ g}
 \mbox{ cm}^{-3}\ll 10^{13}\mbox{ g cm}^{-3}$ and
 temperature $T\lsim 10^{10}\mbox{ K}$, which are typical values
 for the innermost region of a gas shell being ejected.
% {\bf Which region?}
Similar considerations enable us to apply M-B statistics
 for baryons and nuclei, which are obviously farther from
 degeneracy since $m_{\text{nuc}}\gg m_{\text{e}}$, where
 $m_{\text{nuc}}$ is the average mass per nucleon.

We find that kinetic motions of electrons are relativistic
 while that of baryons and nuclei being non-relativistic
 by comparing their rest masses and $k_{\text{B}}T$, namely
\begin{equation}
  m_\text{e}c^2=0.511\mbox{MeV}\lsim k_\text{B}T
  \ll m_\text{p}c^2\simeq 931\mbox{MeV}< m_\text{nuc}c^2\ .
\end{equation}
%here $c$ is the speed of light in vacuum.
%
We first estimate the ratio of one nucleus
 momentum $p_{\text{nuc}}$ to one electron
 momentum $p_{\text{e}}$ as
\begin{equation}
  \dfrac{p_\text{nuc}}{p_\text{e}}
  \sim \dfrac{(2m_\text{nuc}k_\text{B}T)^{1/2} }
  {(k_\text{B}^2T^2/c^2-m_\text{e}^2 c^2)^{1/2}}
  > \bigg( \dfrac{2m_\text{nuc}c^2}{k_\text{B}T} \bigg)^{1/2} \simeq 30\ .
\end{equation}
%
%{\bf Need to check!} {\bf Define notations and avoid
% confusions!}
%{\it
Here $k_\text{B}T$ roughly gives the energy of thermal motion
  of one particle.
The numerical result is based on the assumption that
  the average charge number per nucleus is $2$.
This estimate implies that we can reasonably ignore the diffusion
 effect of electrons when discussing kinetic effect of diffusion:
 electrons are trapped by the electric field when they go inwards,
 but they cannot effectively drag the nuclei in since
 $p_\text{nuc}/p_\text{e}\gg 1$.
Instead, they are dragged by those nuclei; that is,
%to be precise,
 we should discuss $p_\text{nuc}/(Zp_\text{e})$ where $Z$ is the
 average of nuclear charge number, since one nucleus is ``bound''
 with $Z$ electrons on average, yet $p_\text{nuc}/(Zp_\text{e})$
 is still considerably greater than $1$ for a not-very-large $Z$.
Note that the rebound shock has already dissociated a large
 portion of heavy nuclei \citep[e.g.][]{1996A&A...306..167J}.
In other words, diffusion effects on kinetic aspects are
 prevailed by the diffusion process of nuclei rather
 than electrons.
A similar estimation based on the electrostatic
 force and radiation force is presented by e.g.
 \citet{THEORETICAL_ASTROPHYSICS}, yielding similar results.

With a local Cartesian coordinate system erected, with
 the $x-$axis pointing radially outwards, at the contact
 discontinuity interface, probability densities of
 nuclei velocity distribution takes the M-B form of
\begin{equation}
  p(\mathbf{v})\propto\exp{\left\{-\dfrac{\tilde{m}
   [(v_x -v_\text{cd})^2+v_y^2+v_z^2]}{2k_\text{B}T}\right\}}\ ,
\end{equation}
where $v_\text{cd}$ is the radial velocity of the
 contact discontinuity interface and $\tilde{m}$
 is the average mass of one nucleus.
Similar to \citet{springerlink:10.1007/s10509-009-0044-4}, ratio
 $\vartheta$ of the number of particles that diffuses into $r_0$
 from the region $r_0<r<r_0 + l$ ($l$ is the mean free path
 length) during the time period $\delta t$ to the total number
 of particles in that region is approximately given by
\begin{equation}
\begin{split}
  &\vartheta=\left(\int_{r_0}^{r_0+l}4\pi r^2\rho(r)\mbox{d}r
    \int p(\mathbf{v})\mbox{d}^3\mathbf{v}\right)^{-1}
  \\
  &\times\int_{r_0}^{r_0+l}4\pi r^2\rho(r)\mbox{d}r
  \int_{|r+v_x\delta t|<r_0 ,(v_y^2+v_z^2)<(r_0/\delta t)^2}
  p(\mathbf{v})\mbox{d}^3\mathbf{v}\ .
\end{split}
\end{equation}
% Assuming that $\rho(r)$ would not vary very fast inside the contact
% surface and $l$ is not very long,
The mean value theorem for integrals enables us to treat
  $\rho$ and $r^2$ in the integrand as approximate constants,
  especially for those cases with $l/r_0\ll 1$
  [this is guaranteed by estimates presented by eq.
  \eqref{eq:DIFFUSION_VARIABLES_ESTIMATE}], which
  directly leads to
\begin{equation}
\begin{split}
  & \vartheta =\dfrac{1-\exp\Big[-\frac{\tilde{m} r_0^2}
    {2k_\text{B}T(\delta t)^2}\Big]}{\pi^{1/2}}\dfrac{r_0}{l}
  \\
  & \qquad\quad\times\int_1^{1+l/r_0}\mbox{d}\tilde{x}
    \int_{\big[-\frac{r_0}{\delta
      t}(1+\tilde{x})-v_\text{cd}\big]\sqrt{\frac{\tilde{m}}
      {2k_\text{B}T}}}^{\big[\frac{r_0}{\delta t}
  (1-\tilde{x})-v_\text{cd}\big]\sqrt{\frac{\tilde{m}}{2k_\text{B}T}}}
  e^{-\tilde{v}^2}\mbox{d}\tilde{v}\ ,
\end{split}
\end{equation}
where $\tilde{x}=r/r_0$ and $\tilde{v}=v_x[\tilde{m}/
  (2k_\text{B}T)]^{1/2}$ are dimensionless integration
  variables converted from $r$ and $v_x$, respectively.

One sensible choice of $\delta t$ is
 $\delta t=r_0/v_\text{cd}$, reflecting the time scale during
 which the void radius almost doubles ($v_\text{cd}$ is
 time-dependent in polytropic cases, yet this is still an
 indication of radial flow speed for a period of time).
With this choice, we have
\begin{equation}
\begin{split}
  &\vartheta=\dfrac{r_0}{2l}\bigg(1-e^{-\mathcal{V}^2}\bigg)
  \\
  &\qquad\quad\times\int_1^{1+r_0/l}\mbox{d}\tilde{x}
  \big[\text{erf}(2\mathcal{V}+2\tilde{x})
  -\text{erf}(\mathcal{V}\tilde{x})\big]\ ,
\end{split}
\end{equation}
where we define the dimensionless expansion speed
  $\mathcal{V}$ of the contact discontinuity interface
\begin{equation}
  \mathcal{V}=v_\text{cd}
  \bigg(\dfrac{\tilde{m}}{2k_\text{B}T}\bigg)^{1/2}\
\end{equation}
and $\mbox{erf}(x)$ is the standard
 error function of argument $x$.

%{\it
For the scenario of SN remnants, the mean free path $l$ of
 particles near the contact discontinuity interface is
\begin{equation}
  \label{eq:MEAN_FREE_LENGTH_COULOMB}
  l=\dfrac{1}{n\Sigma}\sim\dfrac{m_\text{p}}
  {\rho\Sigma}\ .
\end{equation}
Here $\Sigma$ can be either the cross section of particles in
  Coulomb scattering $\Sigma_\text{cou}$ (scattered by other
  charged particles near the contact discontinuity interface;
  subscript ``cou'' for ``Coulomb'') corresponding to
  $l_\text{cou}$, or Thomson scattering $\Sigma_\text{ph}$
  (scattered by photons inside the bubble; subscript
  ``ph" for ``photon'') corresponding to $l_\text{ph}$.
These two kinds of scatterings are dominant when evaluating the
  diffusion processes of matters near the contact discontinuity
  interface, i.e., particles of those matters can be scattered
  by either other charged particles near the shell inner
  boundary or the photons inside the bubble.

$\Sigma_\text{cou}$ is estimated in \citet[][]{JACKSON_CED}
  by taking that the radius of charge distribution as
  $\sim 1.2\mbox{ f m}\times A^{1/3}$
% eq.(13-104) in the second edition.
\begin{equation}
  \Sigma_\text{cou}\sim 30\mbox{ f m}^2Z^{2/3}
  \left(\dfrac{Z^2e^2}{\hbar v}\right)^2\ ,
\end{equation}
  and here $e$ is the unit electric charge, $Z$ is the average
  charge number per nucleon (assuming $A/Z\simeq 2$ and thus
  $\tilde{m}\simeq 2Zm_\text{p}$), $v$ is to be estimated by
  the magnitude of thermal velocity
  ($v\simeq [k_\text{B}T/(Zm_\text{p})]^{1/2}$), and
  $\hbar=h/(2\pi)$ is the reduced Planck constant.
For $T\lsim 10^{10}\mbox{ K}$ and $Z\geq 1$, we get
  $\Sigma_\text{cou}\gsim 10^{-26}\mbox{ cm}^2$ (this may
  be even higher since
  $\Sigma_\text{cou}\propto Z^{17/3}$) and according to eq. \eqref
  {eq:MEAN_FREE_LENGTH_COULOMB}, $l_\text{cou}\lsim 10^{-7}\mbox{ cm}$.

For the Thomson case of scattering with photons, we have from
  eq. \eqref {eq:THOMSON_SECTION} (for the electron-photon scattering)
  and the number density of photons given by Bose-Einstein statistics
  \citep[e.g.][]{CALLEN_THERMODYNAMICS}:
\begin{equation}
  n_\text{ph}\sim 10^{31}\mbox{ cm}^{-3}\left(
    \dfrac{T_\text{rad}}{10^{10}\mbox{ K}}\right)^3\ ,
\end{equation}
  then the mean free path length $l_\text{ph}$ is also estimated to be
  $l_\text{ph} \sim 10^{-7}\mbox{ cm}$ as well for the radiation in the
  bubble whose temperature is $T_\text{rad}\sim 10^{10}\mbox{ K}$.
%  } {\bf Please clarify}.

  % (note that $\rho$ here is the mass density at the immediately
  % inside of the contact discontinuity, which should not drop so
  % sharply as a cliff in physical reality, in contrast to the ideal
  % model of discontinuity mathematically, see figure 4 of
  % \citealt{1985ApJ...295...14B} for details).

Let us consider a case with these features common
  for SN ejecta when a SN explosion takes place, e.g.
  $r_0\sim 10^8\mbox{ cm}$,
  $v_\text{cd}\sim 10^9\mbox{ cm s}^{-1}$, $\tilde{m}\sim
  4m_\text{p}$, $T_\text{cd}\sim 10^{10}\mbox{ K}$, and
  $\rho_\text{cd}\sim 10^{10}\mbox{ g cm}^{-3}$.
These parameters yield following conditions:
\begin{equation}
  \label{eq:DIFFUSION_VARIABLES_ESTIMATE}
  \mathcal{V}\sim 1.6\ ,\qquad\qquad\quad
  \dfrac{l_\text{cou}}{r_0}
  \sim\dfrac{l_\text{ph}}{r_0}\lsim 10^{-15}\ll 1\ ;
\end{equation}
  thus during a time interval of $\delta t=r_0/v_\text{cd}$,
\begin{equation}
  \vartheta\simeq\dfrac{(1-e^{-\mathcal{V}^2})}{2}
  \Big[\text{erf}(2\mathcal{V}+2)-\text{erf}(\mathcal{V})
  \Big]\sim 1.3\%\ .
\end{equation}
This is such a low ratio of paticle diffusion that
  our void model with sharp discontinuous interface
  is well justified in the context of SNe.

Although this estimation is carried out during a SN explosion, we
  expect it to be hold over a relatively long period of time.
Since $\mathcal{V}\propto v_\text{cd}/T$ and $T\propto p/\rho$
  for an ideal gas, we then have $\mathcal{V}\propto t^0$
  according to eq. \eqref{eq:SELF_SIMILAR_TRANSFORMATION}.
Also, initial values of $l_\text{c}/r_0$ and $l_\text{ph}/r_0$ are
  so small that they will not increase to be comparable with $1$
  quickly.

\section{Gradual Transition of Phase Diagram
  for Solutions Crossing the SCC Smoothly }

\begin{figure}
  \includegraphics[width=80mm]
  {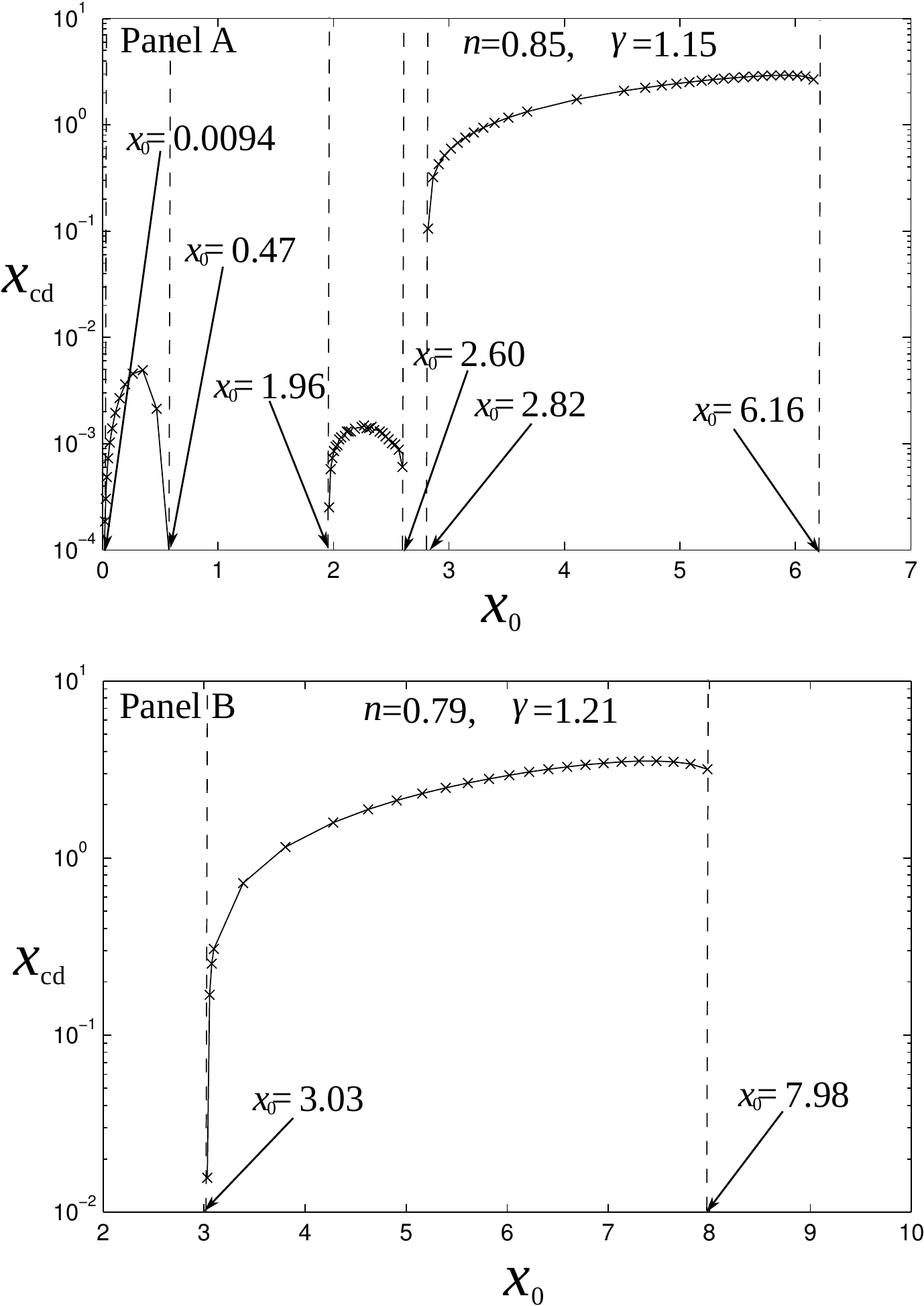}
  \caption{
A direct comparison between relations of $x_\text{cd}$ (where
  $\alpha_\text{cd}$ and $v_\text{cd}$ can be determined)
  versus $x_0$ on the SCC for $n=0.85$ (Panel A) and for
  $n=0.79$ (Panel B).
The light solid curves marked along with crosses (``$\times$")
  represent branches of type-1 void solutions.
In Panel A, type-1 void solutions exist for three intervals
  of $x_0$: $0.0094<x_0<0.47$, $1.96<x_0<2.60$, and $2.82<x_0<6.16$.
In Panel B in contrast, there exists only one branch of type-1
  void solution for $x_0$ in the interval $3.03<x_0<7.98$.
We find by extensive numerical explorations that there are
  actually no type-2 void solutions when $n\lsim 0.87$, while
  the two branches of type-1 void solution on the left and in
  the middle of Panel A also disappear when $n\lsim 0.80$.
  }
  \label{fig:SCC_PD_0_85_0_79}
\end{figure}

Readers might be interested in the gradual variation of phase
  diagrams indicating the transition from Fig. \ref{fig:SCC_PD_0_9}
  to Fig. \ref{fig:SCC_PD_0_67}: e.g., where do some curves go
  in Fig. \ref{fig:SCC_PD_0_9} as the value of $n$ is decreased?
In Appendix E here, we briefly show with the help of Fig.
  \ref{fig:SCC_PD_0_85_0_79}, how Fig. \ref{fig:SCC_PD_0_9}
  becomes Fig. \ref{fig:SCC_PD_0_67} as $n$ decreases.

As $n$ drops from $0.9$, the line marked with circles
  (``$\circ$") in Fig. \ref{fig:SCC_PD_0_9}, corresponding
  to the branch of type-2 solutions, gradually ``shrinks"
  (i.e. the range of $x_0$ for this branch of solutions
  becomes narrower).
When $n\simeq 0.87$, this branch completely
  disappears from the phase diagram.
Panel A in Fig. \ref{fig:SCC_PD_0_85_0_79} shows the
  phase diagram for $n=0.85$ when the branch of type-2
  solution has already disappeared.

When $n$ value is reduced further to $n\simeq 0.805$, the left two
  branches of type-1 solutions, indicated by curves marked with crosses
  (``$\times$") in the phase diagram, disappear almost simultaneously.
That is, only the right-most curve in Panel A of Fig.
  \ref{fig:SCC_PD_0_85_0_79} ``survives".
Panel B in Fig. \ref{fig:SCC_PD_0_85_0_79} shows the phase
  diagram for $n=0.79$ when the left two curves no longer exist.
Qualitatively, there are no significant changes as $n$ continues
  down to $0.67$, except that the range of $x_0$ of this branch of
  solutions expands to the extent shown in Fig. \ref{fig:SCC_PD_0_67}.

%{\bf Not enough references!! Please pay attention to Lou \& Cao
%papers; Hu \& Lou papers; relevant papers cited in Bethe;
%Chevalier papers...}

\bibliographystyle{mn2e}
\bibliography{ThisBib}
%%%%%%%%%%%%%%%%%%%%%%%%%%%%%%%%%%%%%%%%%%%%%%%%%%%%%%%%%%%%%%%%%%%%%%%%%%%%%%%%
%Bibliography part is generated by BibTeX via file ThisBib.bib automatically.  %
%%%%%%%%%%%%%%%%%%%%%%%%%%%%%%%%%%%%%%%%%%%%%%%%%%%%%%%%%%%%%%%%%%%%%%%%%%%%%%%%

\label{lastpage}

\end{document}